\documentclass[twocolumn]{aastex62}
\usepackage{natbib}
\received{May 30, 2019}
\revised{September 11, 2019}
\submitjournal{ApJ}

\shorttitle{Radio emission from SNe in the very early phase}
\shortauthors{Matsuoka et al.}

\begin{document}

\title{Radio Emission from Supernovae in the Very Early Phase: Implications for the Dynamical Mass Loss of Massive Stars}

\correspondingauthor{Tomoki Matsuoka}
\email{t.matsuoka@kusastro.kyoto-u.ac.jp}

\author{Tomoki Matsuoka}
\affil{Department of Astronomy, Kyoto University, \\
Kitashirakawa-Oiwake-cho, Sakyo-ku, Kyoto, 606-8502, Japan}

\author{Keiichi Maeda}
\affiliation{Department of Astronomy, Kyoto University, \\
Kitashirakawa-Oiwake-cho, Sakyo-ku, Kyoto, 606-8502, Japan}

\author{Shiu-Hang Lee}
\affiliation{Department of Astronomy, Kyoto University, \\
Kitashirakawa-Oiwake-cho, Sakyo-ku, Kyoto, 606-8502, Japan}

\author{Haruo Yasuda}
\affiliation{Department of Astronomy, Kyoto University, \\
Kitashirakawa-Oiwake-cho, Sakyo-ku, Kyoto, 606-8502, Japan}

\begin{abstract}
Recent high-cadence transient surveys and rapid follow-up observations indicate that some massive stars may dynamically lose their own mass within decades before supernovae (SNe). Such a mass-loss forms `confined' circumstellar medium (CSM); a high density material distributed only within a small radius ($\lesssim 10^{15}$ cm with the mass-loss rate of 0.01 $\sim 10^{-4} M_\odot$ yr$^{-1}$). While the SN shock should trigger particle acceleration and magnetic field amplification in the `confined' CSM, synchrotron emission may be masked in centimeter wavelengths due to free-free absorption; the millimeter range can however be a potential new window. We investigate the time evolution of synchrotron radiation from the system of a red super giant surrounded by the `confined' CSM, relevant to typical type II-P SNe. We show that synchrotron millimeter emission is generally detectable, and the signal can be used as a sensitive tracer of the nature of the `confined' CSM; it traces different CSM density parameter space than in the optical. Furthermore, our simulations show that the `confined' CSM efficiently produces secondary electrons and positrons through proton inelastic collisions, which can become main contributors to the synchrotron emission in several ten days since the SN. We predict that the synchrotron emission is detectable by ALMA, and suggest that it will provide a robust evidence of the existence of the `confined' CSM.
\end{abstract}


\section{Introduction} \label{sec:intro}
Massive stars with a hydrogen rich envelope such as a red super giant (RSG) are known to end in an explosive phenomenon called Type II supernovae \citep[SNe,][]{2015PASA...32...16S}. Radiation from the SN contains information about the progenitor or its surrounding environment, providing an imprint of the stellar evolution of massive stars \citep{1997ARA&A..35..309F}. Thanks to high-cadence transient surveys and rapid follow-up observations, such as the Palomar Transient Surveys \citep{2009PASP..121.1395L, 2009PASP..121.1334R} and the Zwicky Transient Facility \citep{2019PASP..131a8002B, 2019arXiv190201945G}, it has been observationally suggested that some massive stars release  a large amount of their own mass just before the SN,  potentially in an eruptive way \citep{2014Natur.509..471G, 2016ApJ...818....3K}.

SN 2013fs is one of the objects for which the above picture has been robustly established 
\citep{2017NatPh..13..510Y}.
This is classified as a Type II-P SN by the P-Cygni profile of H$\alpha$ line and the optical light curve. It showed highly ionized lines in spectra taken several hours after the discovery  (`flash spectra'). Based on the H$\alpha$ line luminosity in these spectra and upper limit of non-thermal emission in the late phase, they concluded that the progenitor of SN 2013fs may have dense circumstellar material (CSM) only in a small region ($R_{\rm CSM} \lesssim 10^{15}$ cm), which is named a `confined' CSM. This structure corresponds to the mass-loss rate of $\dot{M} \gtrsim 10^{-4} \ M_\odot$ yr$^{-1}$ for a wind velocity of $u_w = 100$ km s$^{-1}$. This indicates that massive stars, at least some RSGs, release their own mass violently decades before the SN (Figure 1). 
This picture challenges the existing stellar evolution models, and has triggered new activities of revising the theory of the stellar evolution 
\citep[e.g.,][]{2012ARA&A..50..107L, 2017MNRAS.470.1642F, 2019arXiv190407878O}. Light curve analyses also provide the indirect evidence for the dynamical mass-loss \citep[e.g.,][]{2017ApJ...838...28M, 2017MNRAS.469L.108M, 2018NatAs...2..808F}. 

Further progress in this field may however be encountered by several challenges. The flash spectroscopy has been the most robust method, but there are two difficulties. First is the observational challenge. The `confined' CSM is not distributed beyond $\sim 10^{15}$ cm. Hence, the characteristic timescale for the flash spectra is an order of hours ($\sim R_{\rm CSM} c^{-1}$, where $R_{\rm CSM} \sim 10^{15}$ cm and $c$ is the speed of light). Second is the interpretation. It may not be straightforward to accurately evaluate the CSM density through the radiation transfer model \citep{2014A&A...572L..11G}.

In this study, we suggest that radio emission is key to solving these problems. 
Synchrotron radiation is exclusively produced by the interaction between the shockwave and CSM \citep{1982ApJ...259..302C, 2012ApJ...758...81M, 2017hsn..book..875C}. Thus, we expect that the radio emission could provide the robust evidence of the existence of the CSM. In addition, the observational requirement of this radio diagnostic can be quite moderate as compared to the optical spectral diagnostic; the characteristic time scale in this process is $\sim$ a few days ($\sim R_{\rm CSM} V_{\rm sh}^{-1}$ instead of $\sim R_{\rm CSM} c^{-1}$, where $V_{\rm sh}$ is a shock velocity).

In this paper, we simulate the radio emission, especially focusing on the millimeter range, from an SN surrounded by a `confined' CSM. This paper is constructed as follows. The initial setup of the progenitor and the `confined' CSM, and the SN explosion hydrodynamics are described in Section 2. 
Methods for the simulations of particle acceleration and the resulting synchrotron emission are described in Section 3. The results are presented in Section 4. The paper is closed in Section 5 with conclusions, including discussion on the observational prospects with ALMA.

\section{SN Explosion Hydrodynamics} \label{sec:hydro}
\subsection{Initial Setup}
\begin{figure*}[ht!]
\plottwo{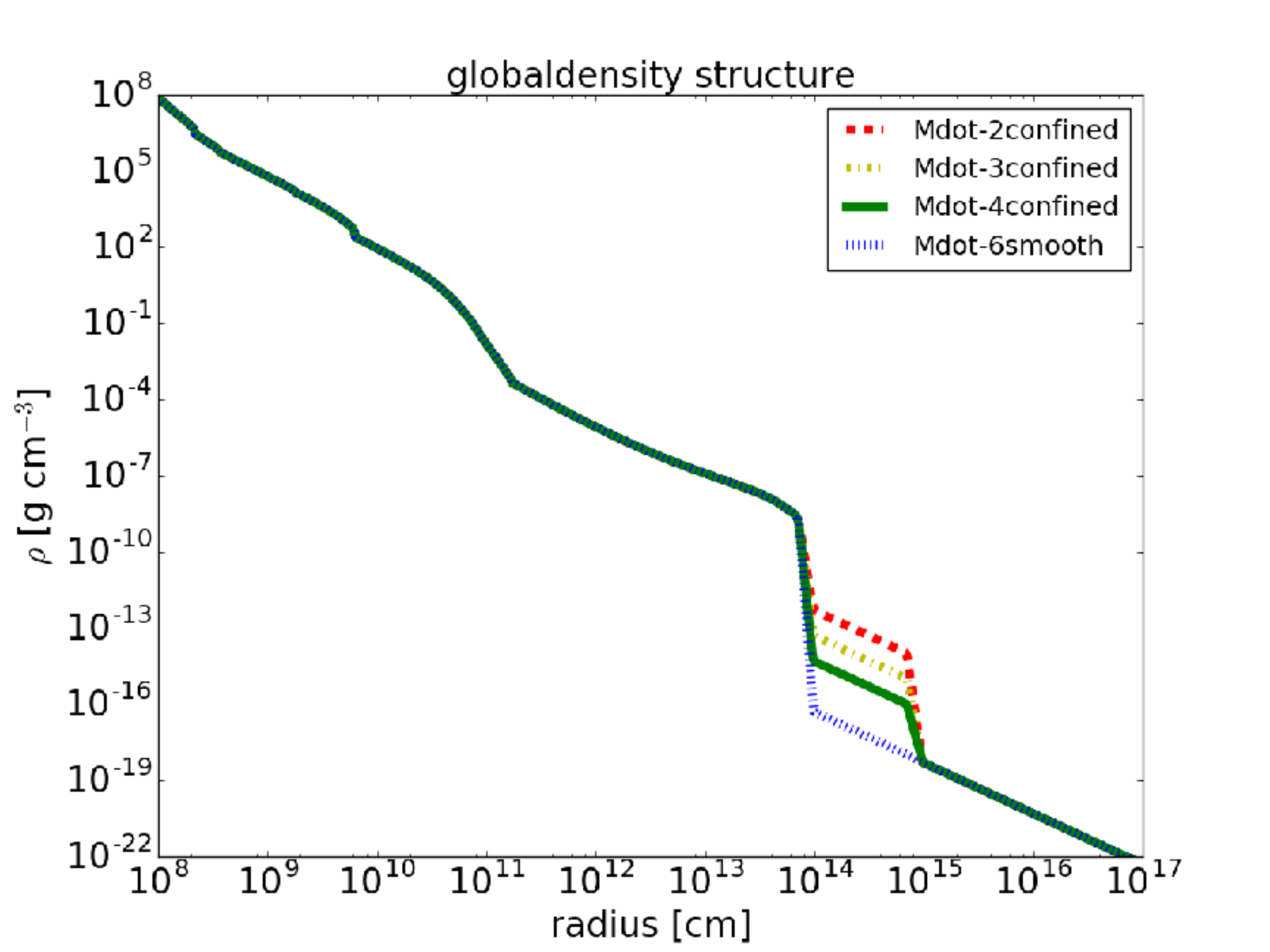}{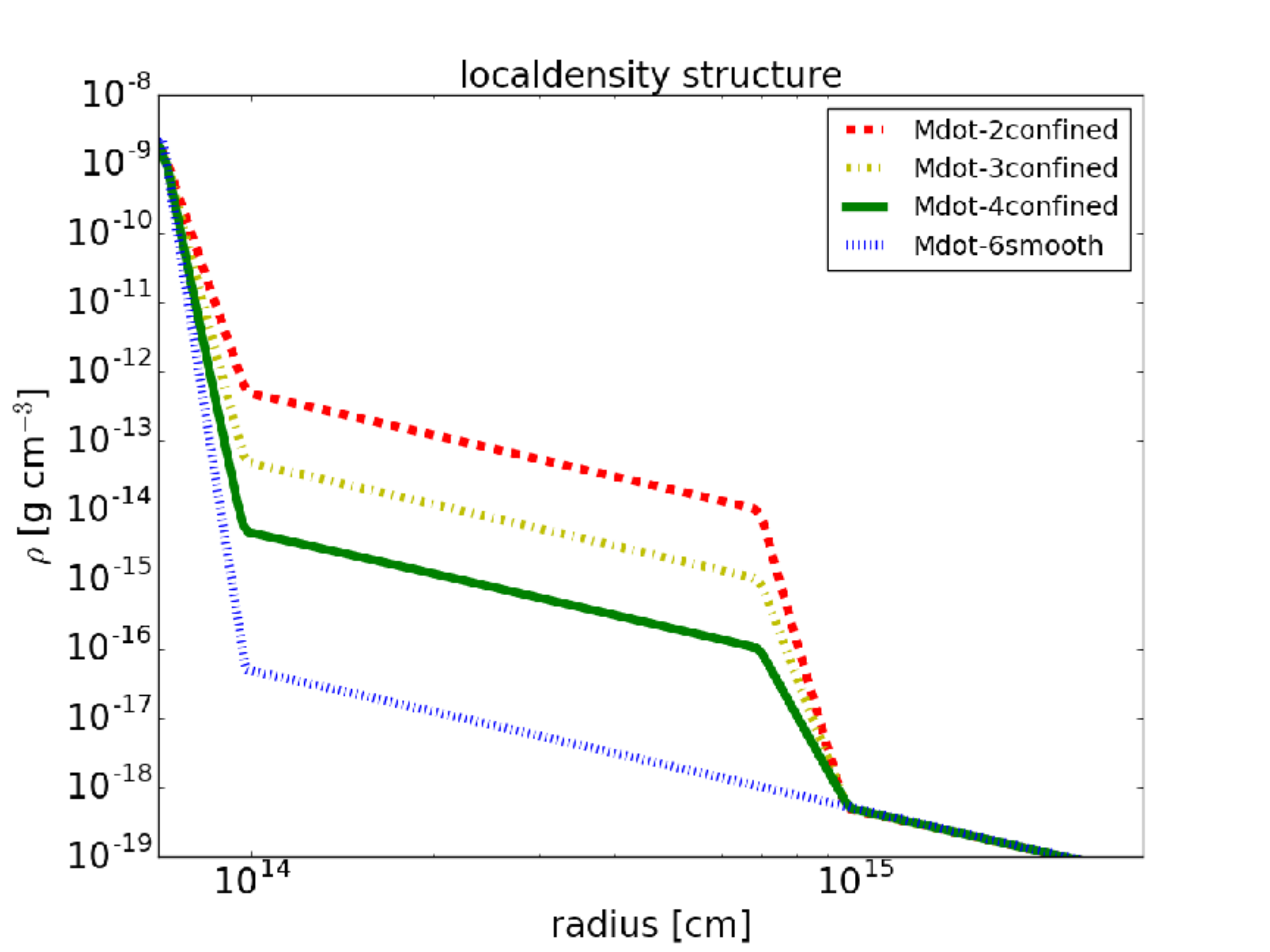}
\caption{Left : The initial density structure of the progenitor and `confined' CSM. The four models adopt the same progenitor profile, but different CSM density. Shown here are Models `Mdot-2confined' (red), `Mdot-3confined' (yellow), `Mdot-4confined' (green), and `Mdot-6smooth' (blue). Right : An expanded view of the `confined' CSM.}
\label{fig:initialCSM}
\end{figure*}
Figure 1 shows the initial density structure of our models. We use an RSG progenitor model which is evolved from a zero-age main sequence star with the initial mass $15M_\odot$, by using the stellar evolution code MESA \citep{2011ApJS..192....3P,2013ApJS..208....4P,2015ApJS..220...15P,2018ApJS..234...34P}. We then attach a `confined' CSM to this progenitor structure by hand, to mimic the observationally inferred distribution \citep{2017NatPh..13..510Y}. 
We examine three `confined' CSM models; high, intermediate, and low density models corresponding to $\dot{M} =10^{-2}, 10^{-3}, 10^{-4}\  M_\odot$ yr$^{-1}$, assuming the wind velocity of $u_{\rm w} = 100$ km s$^{-1}$. The CSM density outside the `confined' CSM is set identical for the three models, which corresponds to $\dot{M} = 10^{-6}\ M_\odot$ yr$^{-1}$ (a typical RSG wind). In addition, a model without the `confined' CSM, which has a smooth RSG wind-type CSM ($\dot{M} = 10^{-6}\ M_\odot$ yr$^{-1}$), is also examined. Hereafter, we name these models `Mdot-2confined', `Mdot-3confined', `Mdot-4confined', and `Mdot-6smooth' (Table 1).

\begin{deluxetable*}{c|ccc}[ht!]
\tablecaption{Models}
\tablehead{
\colhead{model} &
\colhead{Mass-loss rate $[M_\odot$ yr$^{-1}$]}\tablenotemark{a} &
\colhead{CSM mass $[M_\odot]$}\tablenotemark{b} &
\colhead{$R_{\rm CSM}$ [cm]}
}
\startdata
Mdot-2confined&
$10^{-2}$&
0.02&
$7\times 10^{14}$\\
Mdot-3confined&
$10^{-3}$&
$2\times 10^{-3}$&
$7\times 10^{14}$\\
Mdot-4confined&
$10^{-4}$&
$2\times 10^{-4}$&
$7\times 10^{14}$\\
Mdot-6smooth&
$10^{-6}$&
$2\times 10^{-6}$&
no truncation\\
\enddata
\tablenotetext{a}{Assuming $u_{\rm w} = 100$ km s$^{-1}$.}
\tablenotetext{b}{The mass of the CSM within $7\times 10^{14}$ cm.}
\end{deluxetable*}

\subsection{Explosion hydrodynamics}
We employ the 1D Lagrangian spherical symmetric hydrodynamics simulation open code SNEC \citep{2015ApJ...814...63M} and simulate the thermal bomb explosion of the SN for the four models. The explosion energy ($E_{\rm kin}$) and the ejecta mass ($M_{\rm ej}$) are set as $E_{\rm kin} = 10^{51}$ erg and $M_{\rm ej} = 11\ M_\odot$, respectively. 

By following adiabatic hydrodynamics evolution of the system, we trace the position of the shockwave as the shock propagates outward along the Lagrangian coordinate. The simulations provide the information necessary to compute the non-thermal emission properties, e.g., the structure of the shocked CSM. We do not take into account the radiative cooling process such as free-free emission. The radiative cooling timescale is longer than the dynamical timescale in the models `Mdot-3confined', `Mdot-4confined', and `Mdot-6smooth'. The radiative cooling timescale is estimated as follows;
\begin{eqnarray}
t_{\rm cool} &\sim& \frac{5k_{\rm B}T}{n \Lambda_{\rm cool}(T)} \nonumber \\
&\sim& 7 \times 10^5
\left( \frac{n}{10^{10} \ {\rm cm}^{-3}} \right)^{-1}
\left( \frac{T}{10^9 \ {\rm K}} \right)^{0.5} \ {\rm s},
\end{eqnarray}
where $k_{\rm B}, T$ and $n$ are Boltzmann constant, the temperature and the number density at the shock front. $\Lambda_{\rm cool}(T) \sim 1.0\times 10^{-22} (T/10^9 \ {\rm K})^{0.5} \ {\rm erg}\ {\rm s}^{-1}\ {\rm cm}^3$ is the cooling function and for the temperature $T \sim 10^9 \ $K free-free emission is dominant \citep{2017hsn..book..875C}. In the most extreme model `Mdot-2confined', this radiative cooling timescale could be shorter than the dynamical timescale. To further investigate the effect of the radiative cooling on the dynamics we conduct radiation hydrodynamics simulation for the model `Mdot-2confined' by SNEC, noting that this might overestimate the effect since this simulation assumes the full thermalization and the blackbody radiation. As a result we find that the shock velocity is $\sim$ 15 \% lower than that in the adiabatic hydrodynamics simulation, but the other physical properties, such as the density structure, have little difference between these two simulations. Thus, in this study we conduct the adiabatic hydrodynamics simulation as a first approximation even for the `Mdot-2confined' model.

Figure 2 shows the evolution of the forward shock. 
In all the models the forward shock is accelerated at $t\sim 2$ days, which announces the shock breakout from the progenitor surface. In addition, the second acceleration of the forward shock is observed at $t \sim 10$ days for the models with the `confined' CSM. This is due to the steep decline of the `confined' CSM density at $r\sim 10^{15}$ cm. It is seen that in the phase $t \gtrsim 20$ days, the shock velocity in all the models converges to the analytic solution \citep{1982ApJ...258..790C, 1982ApJ...259..302C}, where the shock velocity is described as follows;
\begin{eqnarray}
V_{\rm sh} = 2.2\times 10^4 
\left(
\frac{\dot{M}}{10^{-6}\ M_\odot \mbox{yr}^{-1}}
\right)^{-0.15}
\left(
\frac{u_{\rm w}}{100 \mbox{ km s}^{-1}}
\right)^{0.15} \nonumber \\
\times
\left(
\frac{E_{\rm kin}}{10^{51} \mbox{ erg}}
\right)^{0.425}
\left(
\frac{M_{\rm ej}}{11\ M_\odot}
\right)^{-0.275}
\left(
\frac{t}{10\mbox{ days}}
\right)^{-0.15} \mbox{km s}^{-1}.\nonumber \\
\end{eqnarray}
\begin{figure}
\plotone{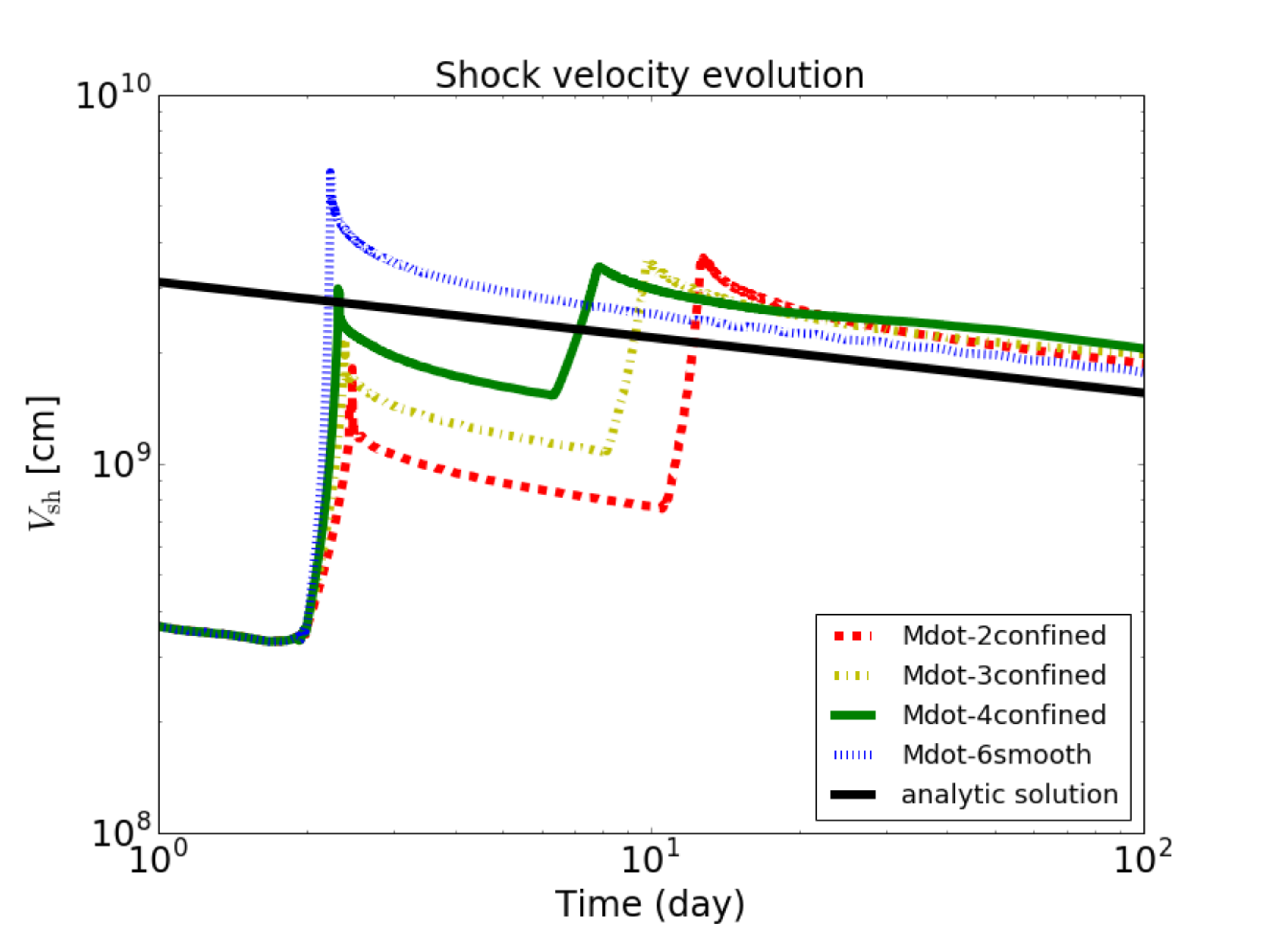}
\caption{The evolution of shockwave velocity as a function of time since the thermal bomb energy injection. Shown here are `Mdot-2confined' (red dashed), `Mdot-3confined' (orange dash-dotteed), `Mdot-4confined' (green solid), and `Mdot-6smooth' (blue dotted). Black line is the analytic solution under thin shell approximation \citep{1982ApJ...259..302C}.} 
\label{fig:velshock}
\end{figure} In this analytic solution, the density distribution of the outer part of the SN ejecta is assumed as $\rho \propto t^{-3} v^{-n}$, where $v$ is the ejecta velocity coordinate. The index is set as $n=8.67$, as obtained by our simulations.

The CSM and ejecta are respectively swept up by the forward shock and reverse shock. These two regions are potential sites for the particle acceleration and synchrotron emission. However, the particle acceleration efficiency at the reverse shock front is believed to be low because the relative velocity of the ejecta in the rest frame of the reverse shock is much lower than that in the forward shock region \citep[see e.g.,][]{2017hsn..book..875C}. In this study, therefore, we neglect the contribution of the reverse shock to the synchrotron emission.


Since we perform adiabatic hydrodynamics simulations, the thermal electron temperature ($T_{\rm e}$) in the unshocked CSM is uncertain. This quantity is indeed important in computing the free-free absorption (FFA) coefficient. In this study, we simply assume $T_{\rm e} = 10^5$ K, a typical value conventionally used in the previous study \citep[see discussion by][]{1998ApJ...509..861F}.

\section{Radio Emission} \label{sec:radio}
Radio emission from SNe is basically identified as synchrotron emission from high-energy non-thermal electrons accelerated in the shocked region \citep{1982ApJ...259..302C}, probably through the diffusive shock acceleration mechanism \citep[DSA;][]{1949PhRv...75.1169F,1978MNRAS.182..147B,1983RPPh...46..973D}. The magnetic field is supposedly amplified either at the shock wave. We consider the non-thermal processes only after the shock breakout, as we will require a collisionless shock. Every time when the shockwave reaches a new Lagrangian zone we compute the magnetic field amplification and the particle acceleration by the DSA using the hydrodynamics information. This Lagrangian zone is now in the shocked region, and we thereafter follow the time evolution of the magnetic field and particle distribution. Here, various cooling processes, as well as the secondary particle (electrons and positrons) injection through hadronic interactions of the shock-accelerated protons, are taken into account. These simulations provide the information for the synchrotron emission, for all the zones between the contact discontinuity and the forward shock. We then solve the radiation transfer for the synchrotron emission along the radial direction.

\subsection{Magnetic field}
In each Lagrangian zone, we compute the `amplified' magnetic field, only once the shock has reached the zone under consideration. This process is modeled as follows, 
\begin{eqnarray}
\frac{B_{\rm sh}^2}{8\pi} = \epsilon_{\rm B} \rho_{\rm sh} V_{\rm sh}^2,
\end{eqnarray}
where $\epsilon_{\rm B}, B_{\rm sh}$ and $\rho_{\rm sh}$ are the amplification efficiency, the amplified magnetic field strength, and the gas density just behind the forward shock. 
With $B_{\rm sh}$ given as the initial condition, the subsequent time evolution of $B(m_r)$ in each zone (where $m_r$ is the mass coordinate), at the shock downstream, is then computed, assuming that the magnetic field flux within the zone is conserved;
\begin{eqnarray}
\frac{d}{dt}
\left[
2\pi r(m_r) \Delta r(m_r) B(m_r)
\right] = 0,
\end{eqnarray}
where $r(m_r)$ is the radius of the zone and $\Delta r(m_r)$ is the thickness of the zone.

\subsection{Particle energy distribution}
Here we consider the energy distribution of the non-thermal particle populations in the shocked region, including electrons, protons, and positrons.
The electrons and protons are assumed to be accelerated through the DSA mechanism (the primary particles). The relativistic protons can further collide with thermal protons
\footnote{In our simulations, we consider only protons for the hadronic interactions. Helium and heavy elements are neglected.}
, producing electrons and positrons through pion decay \citep[the secondary particles;][]{2016MNRAS.460...44P}. Hereafter, we use the subscript i to refer to electrons, protons, or positrons; i = $e^-, p, e^+$.

We denote the number density of the particles i, whose energy is in the range $E_{\rm i} \sim E_{\rm i} + dE_{\rm i}$, as $\frac{dN_{\rm i}}{dE_{\rm i}} dE_{\rm i}$. These densities are traced in all the Lagrangian zones within the shocked CSM region, and obtained by using the following equations;
\begin{eqnarray}
\frac{\partial}{\partial t} 
\left(
\frac{dN_{\rm i}}{dE_{\rm i}}
\right)
= \frac{\partial}{\partial E_{\rm i}} \left( 
\frac{E_{\rm i}}{t_{\rm i, loss}} \frac{dN_{\rm i}}{dE_{\rm i}}
\right)
+ 
\left(
\frac{d \dot{N_{\rm i}}}{d E_{\rm i}}
\right)_{\rm in}, \\
\left(
\frac{d \dot{N_{\rm i}}}{d E_{\rm i}}
\right)_{\rm in}
 = \left(
\frac{d \dot{N_{\rm i}}}{d E_{\rm i}}
\right)_{\rm prim} + 
\left(
\frac{d \dot{N_{\rm i}}}{d E_{\rm i}}
\right)_{\rm sec}.
\end{eqnarray}
The first term in the right hand side of eq. 5 accounts for the cooling processes (for the description of the energy loss timescale $t_{\rm i,loss}$, see Appendix A). For protons, the energy losses by inelastic proton collisions, adiabatic expansion, and coulomb interactions are considered. For the electrons and positrons, synchrotron and inverse Compton (IC) emissions are considered together with the adiabatic expansion and the coulomb interaction as the cooling processes.
In order to determine the IC emission timescale, the seed photon energy density must be specified, and we consider two templates for the bolometric light curve as shown in Figure 3; `default' and `high'. We apply the `default' light curve to all of the CSM models. The effect of the bolometric light curve of seed photons on the radio properties will be discussed in Section 5, where we apply the `high' light curve model to `Mdot-2confined' model. We thereby confirm that our conclusions are insensitive to the choice of the bolometric light curve templates.
\begin{figure}[ht!]
\plotone{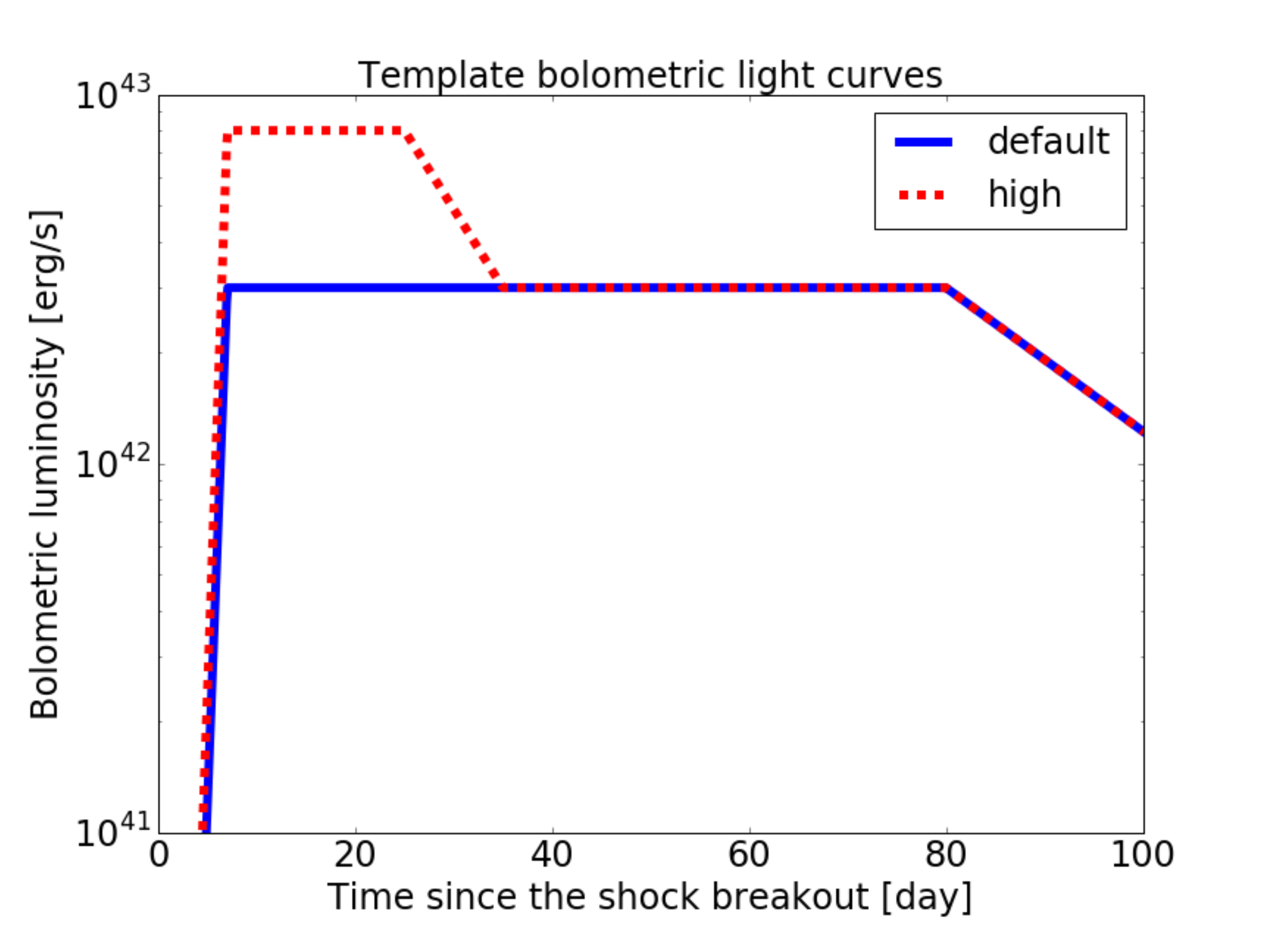}
\caption{The template bolometrc light curve employed in this study (`default' shown by the blue line). For model `Mdot-2confined', we also test the `high' bolometric light curve model (red).}
\label{fig:Lbol}
\end{figure}

The second term on the right hand side in eq. 5 describes the relativistic particle injection. This is divided into two processes (eq. 6). The `primary injection' is for the electrons and protons through the DSA, with the power-law energy distribution. This mechanism is triggered only just behind the forward shock and once per zone. 
The energy distribution is given as follows;
\begin{eqnarray}
\left(
\frac{d \dot{N}_{\rm i}}{dE_{\rm i}}
\right)_{\rm prim}
=\left\{ 
\begin{array}{ll}
C_{\rm i}E_{\rm i}^{-\alpha_{\rm i}}, \ E_{\rm i, min} \leq E_{\rm i} \leq E_{\rm i, max} & \ ({\rm i } = e^-, p) \\
0 & \ ({\rm i } = e^+) \\
\end{array} \right. \nonumber \\
\end{eqnarray}
The normalization coefficient $C_{\rm i}$ is determined by the total energy content of the injected electrons and protons. 
We define the acceleration efficiency $\epsilon_{e^-}$ and $\epsilon_p$ as a fraction of the energy dissipated at the shock transferred to the non-thermal particles;
\begin{eqnarray}
u_{\rm i} = \epsilon_{\rm i} \rho_{\rm sh} V_{\rm sh}^2 \ (\mbox{i} = e^-, p),
\end{eqnarray}
where 
$u_{\rm i}$ is the energy density of the primary injected particles. Then, $C_{\rm i}$ can be estimated by the following integrals,
\begin{eqnarray}
\int\frac{d\dot{N}_{\rm i}}{dE_{\rm i}}dE_{\rm i} &=& {\displaystyle \frac{4\pi R_{\rm sh}^2 n_{\rm i} V_{\rm sh}}{\mathcal{V}}}, \\ 
u_{\rm i} \int\frac{d\dot{N}_{\rm i}}{dE_{\rm i}} dE_{\rm i} &=& 
n_{\rm i} {\displaystyle \int E_{\rm i} \frac{d\dot{N}_{\rm i}}{dE_{\rm i}} dE_{\rm i}} \ (\text{i = } e^-, p),
\end{eqnarray}
where $\mathcal{V}$ is the volume of the zone at the shock front.

The maximum energy $E_{\rm i,max}$ is computed by imposing three physical criteria. The first one is set by the requirement that the acceleration timescale $(t_{\rm i, acc})$ cannot be longer than the energy loss timescale $(t_{\rm i,loss})$. Thus the maximum enegy is limited by
\begin{eqnarray}
t_{\rm i, acc} = t_{\rm i,loss} \ ({\rm i} = e^-, p).
\end{eqnarray}
The acceleration timescale is given as follows \citep{1983RPPh...46..973D};
\begin{eqnarray}
t_{\rm i, acc} = \frac{20}{3}\frac{c^2}{V_{\rm sh}^2} \frac{E_{\rm i}}{qB_{\rm sh}c} \ ({\rm i} = e^-, p).
\end{eqnarray}
Here, $q$ is the elementary charge. 
The second limit comes from the current SN age which is the maximum time available for the acceleration of charged particles since the SN shock breakout. 
Therefore, we impose the constrain on the maximum energy as follows;
\begin{eqnarray}
t_{\rm i, acc} = t - t_{\rm sbo} \ ({\rm i} = e^-, p),
\end{eqnarray}
where $t_{\rm sbo}$ is the time at the shock breakout. 
Finally, escape of the accelerated particles also limits $E_{\rm i,max}$, since the diffusion length is longer for particles with higher energy. 
We impose this constrain on the maximum energy by equating the diffusion length ($L_{\rm diff}$) to the separation between the shock and the upstream free-escape-boundary \citep[$L_{\rm feb}$,][]{2009MNRAS.396.2065C,2012ApJ...750..156L,2019ApJ...876...27Y} as follows;
\begin{eqnarray}
L_{\rm diff} &=& L_{\rm feb}, \\
L_{\rm diff} &=& \frac{\eta}{3}\frac{c}{V_{\rm sh}}\frac{E_{\rm i}}{qB} \ ({\rm i} = e^-, p),\\
L_{\rm feb} &=& f_{\rm esc} R_{\rm sh},
\end{eqnarray}
where $\eta$ is related to nature of the magnetic field turbulence. Here we assume the Bohm limit $\eta = 1$ \citep[see e.g.,][]{1983RPPh...46..973D}. $L_{\rm feb}$ is parametrized by $f_{\rm esc}$, and we set $f_{\rm esc} = 0.1$, which is a typical value inferred from recent models of young supernova remnants such as Tycho \citep{2014ApJ...783...33S} and Vela Jr. \citep{2013ApJ...767...20L}. The maximum energy $E_{\rm i,max}$ is set as the smallest value among those obtained by the three criteria. 
We note that the maximum energy is regulated dominantly by the synchrotron cooling for the electrons, and the escape limit for the protons, in the dense `confined' CSM.

The minimum energy is fixed as $E_{\rm i, min} = 2.5 m_{\rm i} c^2$ where $m_{\rm i}$ is the mass of the particle i. Finally, the power-law index $\alpha_{\rm i}$ is chosen as 3. This is commonly used to explain the radio emission from SNe \citep{2006ApJ...651..381C, 2006ApJ...641.1029C, 2013ApJ...762...14M}.

The second mechanism of the high energy particle injection is through pion decay, triggered by proton inelastic collisions. We focus on the reaction path where electrons and positrons are generated. In this paper, we call these electrons and positrons `secondary particles'. These secondary particles are produced not only at the forward shock but also in all the zones within the shocked region; the accelerated protons remain energetic not only just behind the shock but also in the downstream, and keep producing the secondary particles through pion decay. The injection rate of the secondary particles is written as follows;
\begin{eqnarray}
\left(
\frac{d\dot{N}_{\rm i}}{dE_{\rm i}}
\right)_{\rm sec}
=\left\{ 
\begin{array}{ll}
n_{\rm H} c {\displaystyle \int dE_p \frac{dN_p}{dE_p} \frac{d\sigma_{{\rm p},2}(E_p, E_{\rm i})}{dE_{\rm i}}} & \ ({\rm i } = e^-, e^+) \\
0 & \ ({\rm i } = p), \\
\end{array} \right. \nonumber \\
\end{eqnarray}
where $n_{\rm H}$ is the number density of target protons. ${d\sigma_{p,2}(E_p, E_{\rm i})}/{dE_{\rm i}}$ is the inclusive cross section of p-p collision that produces the secondary particles having an energy $E_{\rm i}$.
The fitting formulae for the proton inelastic collision cross section have been developed by some previous works based on experiments or Monte Carlo simulations. The formalization by \cite{2006ApJ...647..692K} is frequently used in computing the proton collision in SN remnants. In the present simulation, however, the maximum energy of protons can reach to PeV, 
\footnote{Note that these PeV protons lose their energy before the escape from the SN system.}
which is beyond the applicable energy range of the parametrized formulae in \cite{2006ApJ...647..692K}. The formalism by \cite{2006PhRvD..74c4018K} is applicable to  protons more energetic than 0.1 TeV. In this study we adopt these two formalizations in a hybrid way. 

Free parameters in our simulations are the following; $\epsilon_{\rm B}, \epsilon_{\rm e}, \epsilon_{\rm p}, \alpha_{e^-}, \alpha_p,$ and $E_{\rm i, min}$. These parameters are in principle determined by the shock acceleration physics, but there still remains debate on typical values which represent the real situations \citep{2015ApJ...798L..28C}. When we compare our calculations to the observational data in future works, these parameters will be constrained and used as feedback to first principle calculations. However, in this study we will focus on the general characteristics of the radio light curves instead of matching any particular observation, and thus simply fix them as follows; $\epsilon_{\rm B}=0.1, \epsilon_{\rm e}=0.1, \epsilon_{\rm p}=0.1, \alpha_{e^-}=3, \alpha_p = 3$ and $E_{\rm i, min} = 2.5 m_{\rm i} c^2$.

\subsection{Synchrotron emission and absorption processes}
Based on the energy distribution of electrons and positrons, the synchrotron emission from the shocked CSM region is calculated. The synchrotron intensity $I_\nu$ is computed by integrating the following one-dimensional radiative transfer equation from the CD to infinity,
\begin{eqnarray}
\frac{dI_\nu}{dr} = -\alpha_\nu I_\nu + j_\nu.
\end{eqnarray}
The synchrotron emissivity ($j_\nu$) and synchrotron self-absorption (SSA) coefficient ($\alpha_{\nu, {\rm SSA}}$) are defined as follows;
\begin{eqnarray}
j_{\nu} &=& \frac{1}{4\pi}
\sum_{{\rm i} = e^-,e^+}
\int
P_{\nu}(E_{\rm i})\frac{dN_{\rm i}}{dE_{\rm i}} dE_{\rm i}, \\
\alpha_{\nu,{\rm SSA}} &=& \frac{c^2}{8\pi \nu^2}
\sum_{{\rm i} = e^-,e^+}
\int
\frac{\partial}{\partial E_{\rm i}}\left[ E_{\rm i}^2 P_{\nu}(E_{\rm i}) \right] \frac{1}{E_{\rm i}^2} \frac{dN_{\rm i}}{dE_{\rm i}} dE_{\rm i}, \nonumber \\
\end{eqnarray}
in all the zones in the shocked CSM region \citep{1979rpa..book.....R}. We note that we include the term for positrons. $P_{\nu}(E)$ is the synchrotron radiation power per unit frequency (see Appendix B). In the preshock CSM region, free-free absorption (FFA) is important. FFA coefficient is given as follows;
\begin{eqnarray}
\alpha_{\nu, {\rm ff}} = 0.018 T_{\rm e}^{-3/2} Z^2 n_{\rm e} n_{\rm i} \nu^{-2} \tilde{g}_{\rm ff} \mbox{ cm}^{-1},
\end{eqnarray}
where $T_{e}, Z,$ and $\tilde{g}_{\rm ff}$ are thermal electron temperature, the charge of thermal ions, and the free-free Gaunt factor, and the variables in this equation are given in cgs unit. We assume $T_{\rm e} = 10^5$ K (see Section 2.2). For the Gaunt factor we employ the value given by \cite{1979rpa..book.....R}.

\section{Results} \label{sec:results}
\subsection{Properties of the synchrotron emission}
Figure 4 shows the centimeter light curves for 4 models, at the frequency 22 GHz, frequently employed in radio observations of SNe.
\begin{figure}[ht!]
\plotone{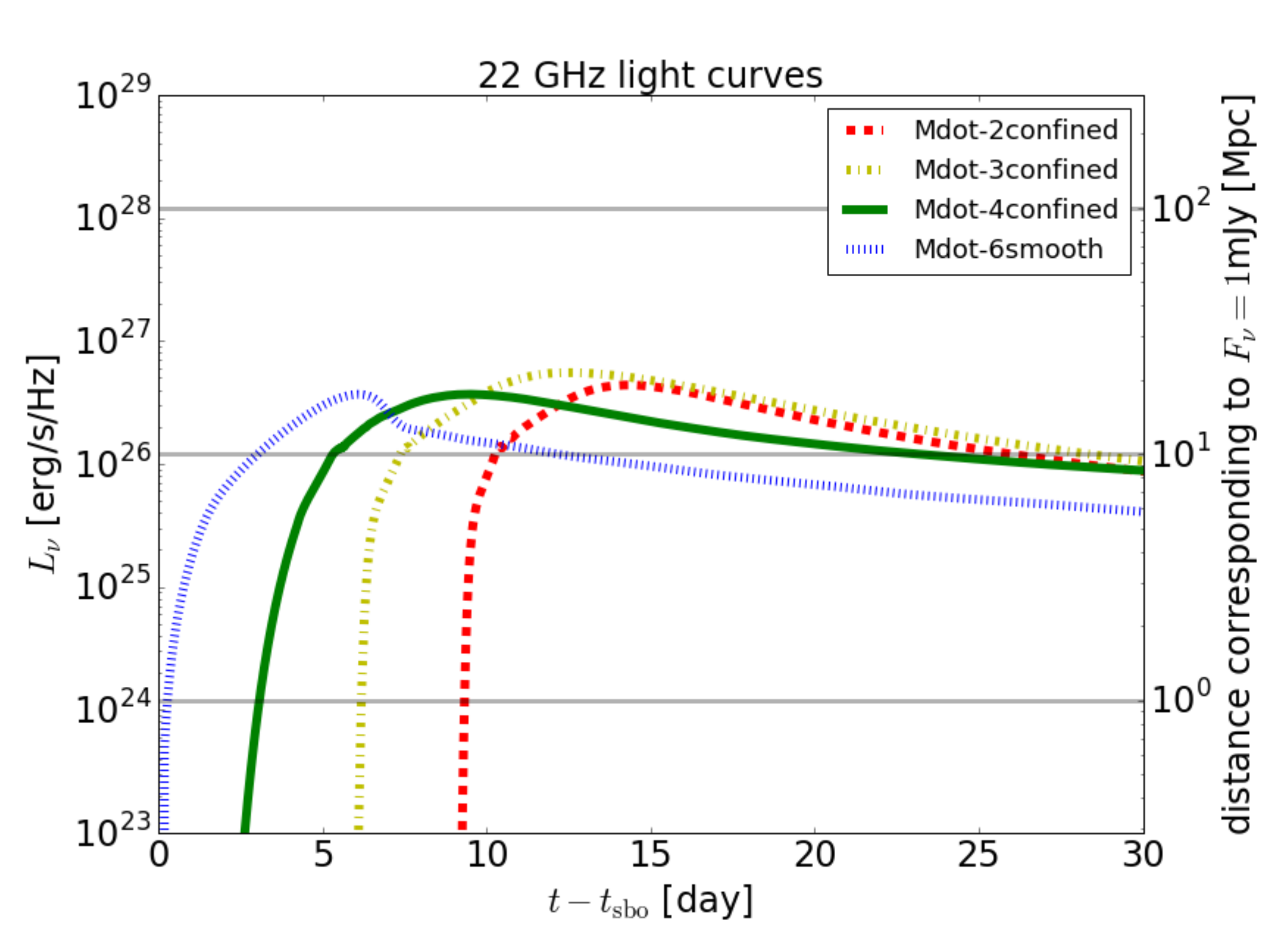}
\caption{The simulated 22 GHz light curves. The x axis is the time since the shock breakout. The y axis expresses the luminosity per unit frequency (left), while the limiting distance where the radio emission is observed with the flux density of 1 mJy is also shown (right). The grey horizontal line shows the corresponding distance at 1, 10, and 100 Mpc.}
\label{fig:centi}
\end{figure}
Our simulations show that the centimeter emission in the first 10 days is weaker for the `confined' CSM with higher density. The synchrotron centimeter emission is damped by SSA or FFA. It will lead to an observational challenge; without robust detection, a large fraction of the information about the nature of the CSM is lost. One will therefore need to go for the higher frequencies, e.g., millimeter, to provide a useful tracer of the `confined' CSM. 

Figure 5 shows the millimeter light curves for 4 models at 100 and 250 GHz. 
\begin{figure}[ht!]
\epsscale{2.0}
\plottwo{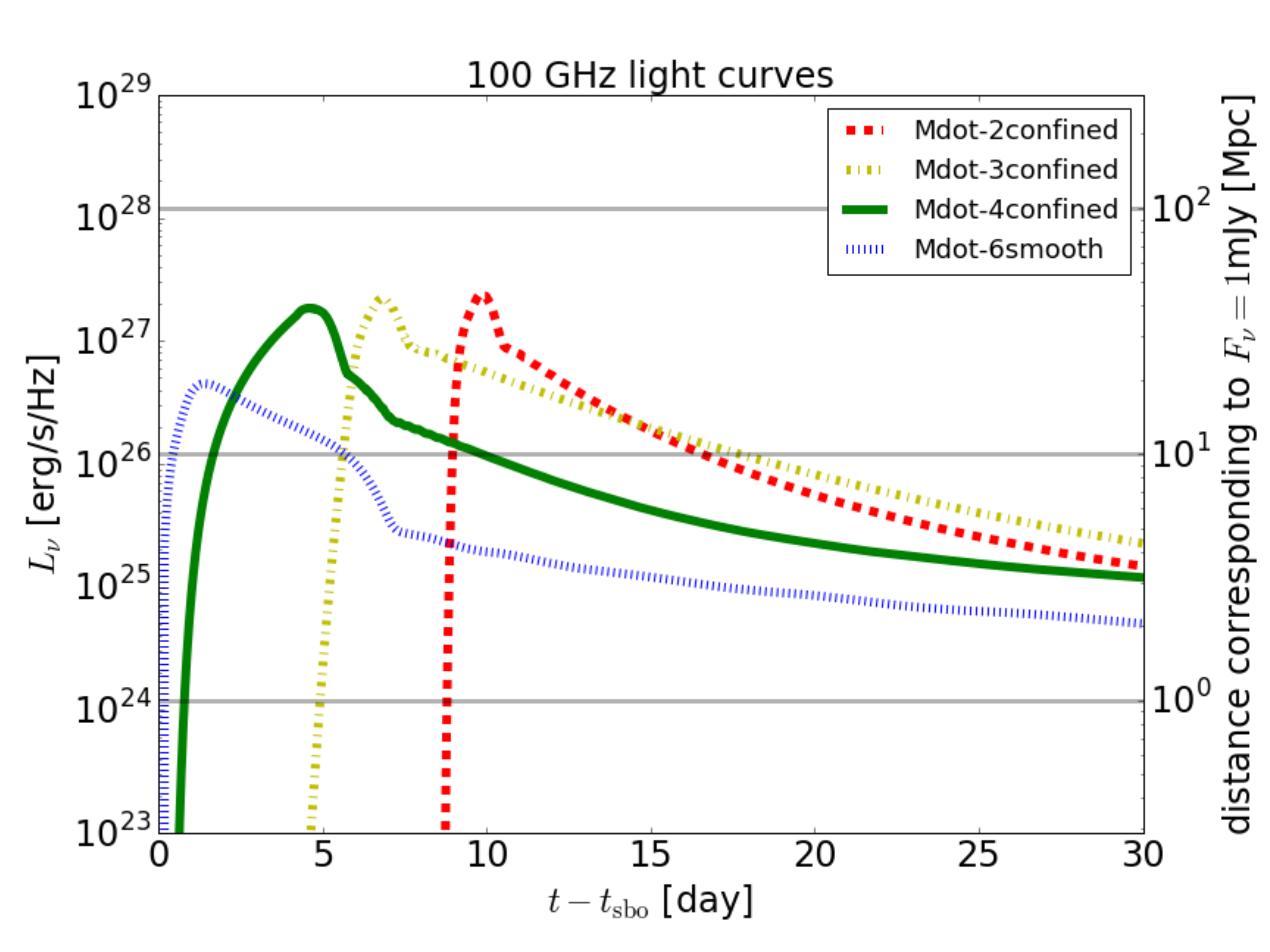}{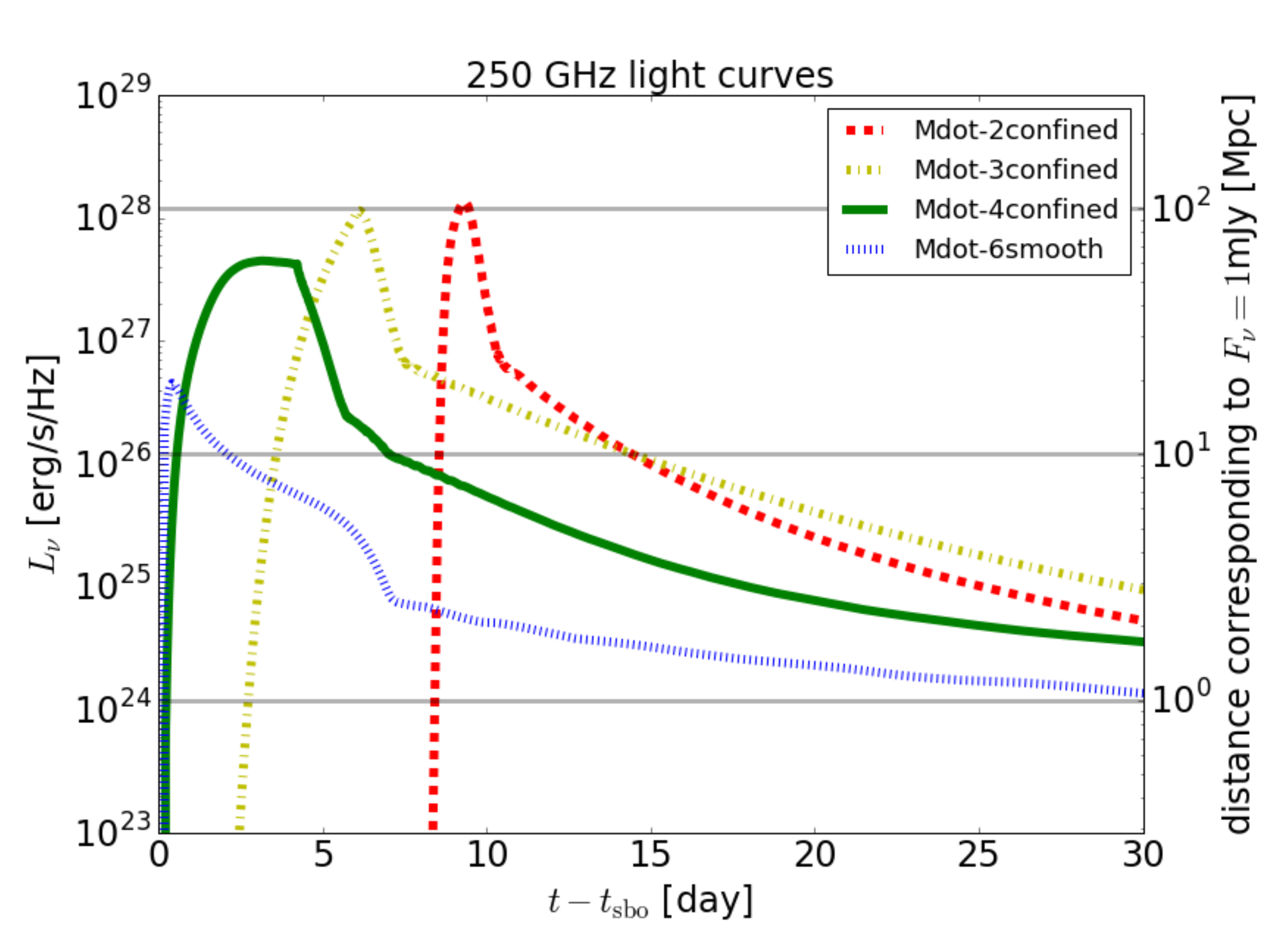}
\caption{The same as Figure 4, but for the millimeter range (100 GHz and 250 GHz).}
\label{fig:milli}
\end{figure}
The peak luminosity in the millimeter emission is enhanced with the `confined' CSM. The shape of the light curve shows different behavior depending on the CSM density.
It is the `Mdot-4confined' model that gives a robustly detectable sign of the millimeter emission in the first 10 days since the shock breakout. The `Mdot-2confined' model shows the millimeter light curve damped in the first 10 days, similar to the case for the centimeter emission. An important difference for the centimeter is the fast variation around $t\sim 10$ days. This phenomenon is caused by FFA in the preshocked CSM region. If the CSM density is low or the synchrotron frequency is high, FFA becomes negligible before the shockwave fully sweeps up the `confined' CSM. In the opposite situation, the synchrotron emission is masked by FFA. When the shock reaches to the outer edge of the `confined' CSM, the synchrotron emission suddenly becomes optically thin, and thus we can observe the millimeter as a fast transient whose variation timescale is a few days. We can analytically investigate this behavior as follows, assuming that FFA is effective only in the `confined' CSM,
\begin{eqnarray}
\int_{R_{\rm sh}}^{R_{\rm CSM}} \alpha_{\nu,{\rm ff}} dr = 1
\Leftrightarrow R_{\rm sh}(\tau_{\rm ff}=1) = R_{\rm CSM} \left\{
1+\mathcal{S}
\right\}^{-1/3}. \nonumber \\
\end{eqnarray}
The characteristic quantity, $\mathcal{S}$, depends on the nature of the CSM and the synchrotron frequency as follows;
\begin{eqnarray}
\mathcal{S} \sim &
10^{-3} {\displaystyle \left(\frac{\dot{M}}{10^{-2}\ M_\odot\mbox{/yr}}\right)^{-2}
\left(\frac{\nu}{100\mbox{ GHz}}\right)^{2}
\left(\frac{R_{\rm CSM}}{7\times 10^{14}\mbox{cm}}\right)^{3}}.\nonumber \\
\end{eqnarray}
The dense CSM or the low frequency leads to substantial FFA and delays the emergence of the synchrotron signal. If the `confined' CSM is opaque to the synchrotron emission, a fast transient-like variant is expected due to the rapid decrease of the optical depth as the shock approaches to the outer edge of the `confined' CSM. In fact in `Mdot-2confined' model $\mathcal{S}$ is smaller than unity and $R_{\rm sh}(\tau_{\rm ff}=1) \sim R_{\rm CSM}$ is realized. On the other hand, if the density of the `confined' CSM is not to much high (e.g., `Mdot-4confined' model at 250 GHz) then $R_{\rm sh}(\tau_{\rm ff}=1) < R_{\rm CSM}$; in this case, the system becomes transparent for FFA when the shock is still propagating in the `confined' CSM. These analyses are consistent with our numerical results.

%
Figure 6 shows the time evolution of SSA and FFA optical depths since the shock breakout.
\begin{figure*}[ht!]
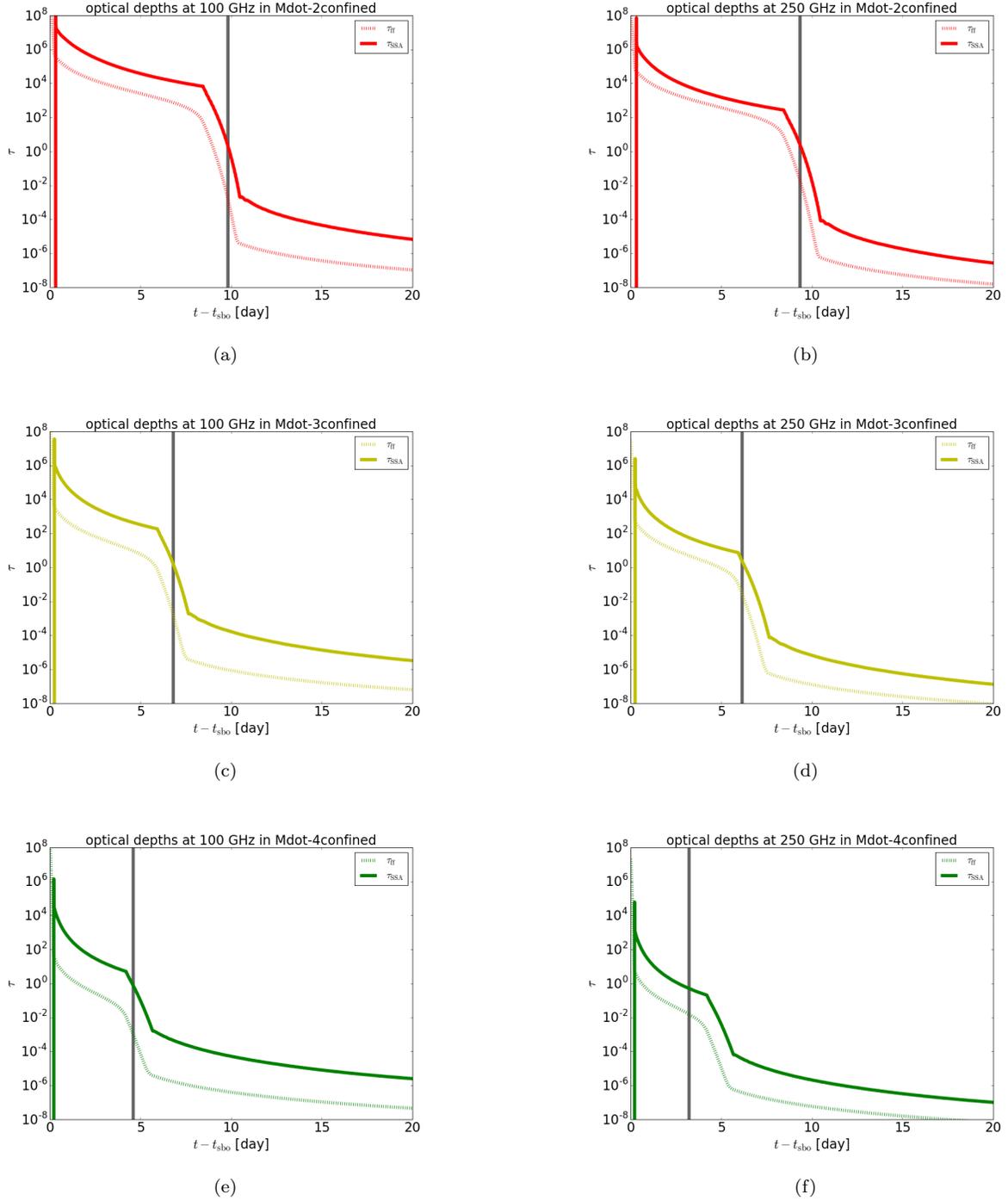

\gridline{
\fig{fig6_a.eps}{0.4\textwidth}{(a)}
\fig{fig6_b.eps}{0.4\textwidth}{(b)}
}
\gridline{
\fig{fig6_c.eps}{0.4\textwidth}{(c)}
\fig{fig6_d.eps}{0.4\textwidth}{(d)}
}
\gridline{
\fig{fig6_e.eps}{0.4\textwidth}{(e)}
\fig{fig6_f.eps}{0.4\textwidth}{(f)}
}
\caption{Time evolutions of the optical depths; FFA (dotted) and SSA (solid). The top, middle, and the bottom panels are for the different `confined' CSM models ($\dot{M} = 10^{-2}, 10^{-3}, 10^{-4} \ M_\odot$ yr$^{-1}$ for the `confined' CSM density). The left and right panels are for 100 GHz and 250 GHz, respectively. The grey vertical line shows the peak date in each models.}
\end{figure*}
We can derive the dependences of these absorption coefficients on the density and frequency as follows;
\begin{eqnarray}
\tau_{\rm ff} & \propto &\rho^2 \nu^{-2}, \\
\tau_{\rm SSA} & \propto & \epsilon_{\rm e} \rho B_{\rm sh}^{(\alpha_{e^-}+2)/2} \nu^{-(\alpha_{e^-}+4)/2} \nonumber \\
&\propto& \epsilon_{\rm e} \epsilon_{\rm B}^{5/4} \rho^{9/4} \nu^{-7/2} \ (\alpha_{e^-} = 3).
\end{eqnarray}
We used the relation $B_{\rm sh} \propto (\epsilon_{\rm B} \rho_{\rm sh})^{1/2}$.
The two absorption processes indeed have similar dependence on the density. Therefore, the difference in the mass-loss rate has little effect on the ratio of the optical depths of the two processes. In our models, $\tau_{\rm SSA}/\tau_{\rm ff} \sim 100$ is realized. We however note that $\tau_{\rm SSA}$ is sensitive to the shock acceleration parameters ($\epsilon_{\rm e}$ and $\epsilon_{\rm B}$). The smaller value of $\epsilon_{\rm e}$ or $\epsilon_{\rm B}$ results in smaller $\tau_{\rm SSA}$. If $\epsilon_{\rm e} \epsilon_{\rm B}^{5/4} \sim 10^{-4}$, then $\tau_{\rm SSA} \sim \tau_{\rm ff}$ is realized, and then the light curve may be exclusively shaped by FFA. 

In figure 6, the grey lines show the peak date, and imply that the maximum luminosity is determined by SSA (but see below for the important role of FFA). 
This can be explained in the same way as the previous SN radio emission studies \citep[e.g., see][]{1998ApJ...499..810C, 2017hsn..book..875C}. The peak luminosity is given by
\begin{eqnarray}
L_{\nu,{\rm peak}} = 4\pi R_{\rm sh}^2 \pi S_{\nu}, 
\end{eqnarray}
where $S_\nu$ is the source function of synchrotron defined as
\begin{eqnarray}
S_{\nu, \rm syn} &=& \frac{8\pi m_e}{\alpha_{e^-}+1}
\left(
\frac{2\pi m_{\rm e}c}{3qB\sin\theta}
\right)^{1/2} \nu^{5/2} \nonumber \\
&\times&
\frac{\Gamma(\alpha_{e^-}/4+19/12)\Gamma(\alpha_{e^-}/4-1/12)}{\Gamma(\alpha_{e^-}/4+1/6)\Gamma(\alpha_{e^-}/4+11/6)}
\end{eqnarray}
\citep{1979rpa..book.....R}. At the peak date the optical depth must be unity,
\begin{eqnarray}
\alpha_{\rm SSA} \Delta R_{\rm sh} = 1,
\end{eqnarray}
where $\Delta R_{\rm sh}$ is the thickness of the shocked region.
Under the standard assumption that the peak luminosity is mostly dominated by the primary electrons (with $\alpha_{e^-}=3$) as confirmed by the simulations, the constraint on the magnetic field is derived as follows;
\begin{eqnarray}
B \propto \nu^{7/9}R_{\rm sh}^{-2/9}.
\end{eqnarray}
The relation between the luminosity and the shock radius is thus expressed as follows;
\begin{eqnarray}
L_{\nu,{\rm peak}} \propto (R_{\rm sh}\nu)^{19/9}.
\end{eqnarray}
The shock radius thus determines the peak luminosity. Both `Mdot-2confined' and `Mdot-3confined' models have similar radius at the peak date, and thus show the similar peak luminosities.

Finally, we will comment on why FFA is important in shaping the light curves even if $\tau_{\rm SSA} \gg \tau_{\rm ff}$. This is explained by the dependence of the intensity on these optical depths; $I_\nu \simeq S_\nu(1-e^{-\tau_{\rm SSA}})e^{-\tau_{\rm ff}}$. The absorber of SSA, a high energy electron, is also the emitter. Thus the large optical depth of SSA leads to the intensity approaching to the source function. The absorber of FFA, on the other hand, is the thermal particles in the preshocked CSM, working as an external absorber. Hence, if $\tau_{\rm ff} \gg 1$ is satisfied, the radio emission is completely damped, unlike SSA. This feature is apparent especially in the first 10 days since the shock breakout, $t \lesssim 10$ days.

\subsection{A role of the secondary particles}
What kind of particles dominates the synchrotron emission is also an interesting question. Figure 7 shows the time evolution of the luminosity in the 4 models, divided into contributions from different sources.
\begin{figure*}[ht!]
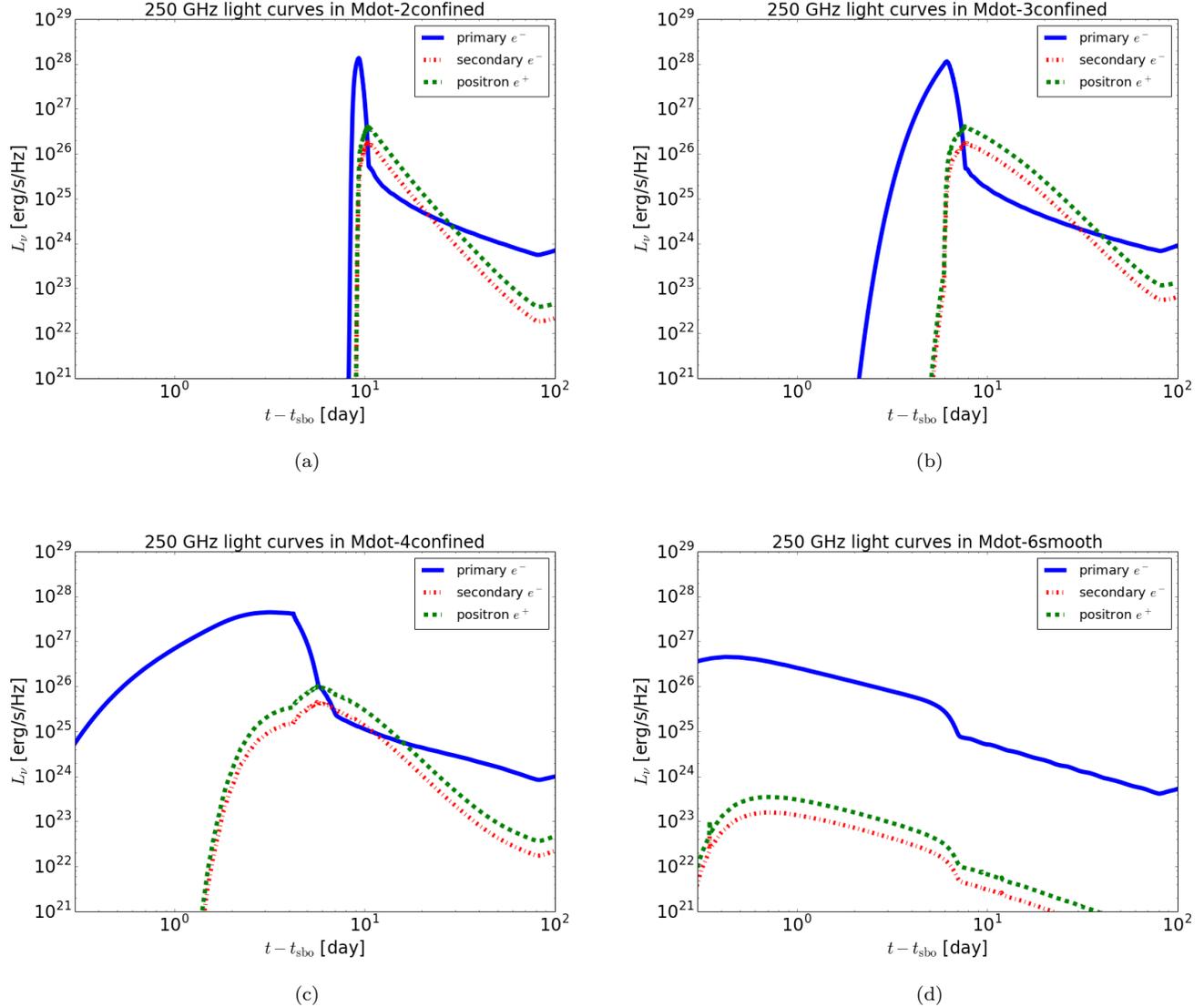

\gridline{
\fig{fig7_a.eps}{0.5\textwidth}{(a)}
\fig{fig7_b.eps}{0.5\textwidth}{(b)}
}
\gridline{
\fig{fig7_c.eps}{0.5\textwidth}{(c)}
\fig{fig7_d.eps}{0.5\textwidth}{(d)}
}
\caption{Light curves for the 4 models, divided into individual contributions by different sources; primary, secondary electrons, and positrons. (a) `Mdot-2confined', (b) `Mdot-3confined', (c) `Mdot-4confined', and (d) `Mdot-6smooth'.}
\label{fig:LCparticle}
\end{figure*}
The emission at peak luminosity is dominated by the primary electrons. For the models with the `confined' CSM, the secondary particles make a large contribution to the synchrotron emission in the phase after the maximum. Especially in `Mdot-2confined' and `Mdot-3confined' models, they overshadow the emission from the primary particles. This is a phenomenon attributed to the existence of the `confined' CSM \citep[see][for a general model with a dense CSM, developed for SNe IIn]{2019ApJ...874...80M}. The protons in the `confined' CSM obtain relativistic energies by the shock acceleration, but their cooling timescale is much longer than those of electrons and positrons. Thus for a while after the propagation of the shockwave in the `confined' CSM, the protons remain energetic and later generate the secondary electrons and positrons.

We note that the time dependence of the luminosity emitted by the secondary particles is steeper than that of the primary particles. This is because the particle injection rate, corresponding to the second term of the right hand side in eq. 6, has different dependence on the CSM density compared to the primary particles. In case of the primary electrons it is $\propto \rho$ \citep[see e.g.,][]{2006ApJ...641.1029C}, while it is $\propto \rho^2$ for the secondary particles. Therefore, the secondary injection is more sensitive to the CSM density and hence decays faster with time than the primary injection.

\section{Conclusions and discussion} 
\label{sec:discussion}
\subsection{Dependence of the inverse Compton cooling on the target photon density}
As mentioned in Section 3.2, the inverse Compton cooling is computed by using the bolometric luminosity $L_{\rm bol}$. Type II-P SNe have a variety in the bolometric light curves, but they generally show a plateau lasting for $\sim$ 100 days since the explosion \citep{2014ApJ...786...67A}. However, there is a possibility that the extremely dense and `confined' CSM may alter the bolometric light curve in the earliest phase through the interaction between the ejecta and CSM \citep[see e.g.,][]{2017ApJ...838...28M}.

For `Mdot-2confined' model, we have tested this possibility by replacing the bolometric light curve (the `high' model in Figure 3). This is motivated by the following estimate of the optical depth $\tau$ in the initial profile of the CSM to the SN thermal photons,
\begin{eqnarray}
\tau &=& \int_{\rm CSM} \kappa \rho dr  \nonumber \\
&\simeq& 10
\left(\frac{\kappa}{0.2 \mbox{ cm$^2$/g}}\right)
\left(\frac{\dot{M}}{10^{-2} M_\odot \mbox{yr} ^{-1}}\right)
\left(\frac{u_{\rm w}}{100\mbox{ km s}^{-1}}\right)^{-1}, \nonumber \\
\end{eqnarray}
where the opacity $\kappa$ is assumed to be coming from electron scatterings. We see that only `Mdot-2confined' model may change the bolometric light curve due to the large optical depth. Figure 8 shows that the radio luminosity after the peak date is decreased by a factor of a few. In the earliest phase, 10 days since the shock breakout, it has little effect. This is due to the intense magnetic field in the shocked CSM; the particle cooling time is determined by synchrotron, not by IC cooling. Therefore, if we focus on the observation around the peak luminosity time, the details of the bolometric luminosity evolution do not affect our conclusions.
\begin{figure}[ht!]
\plotone{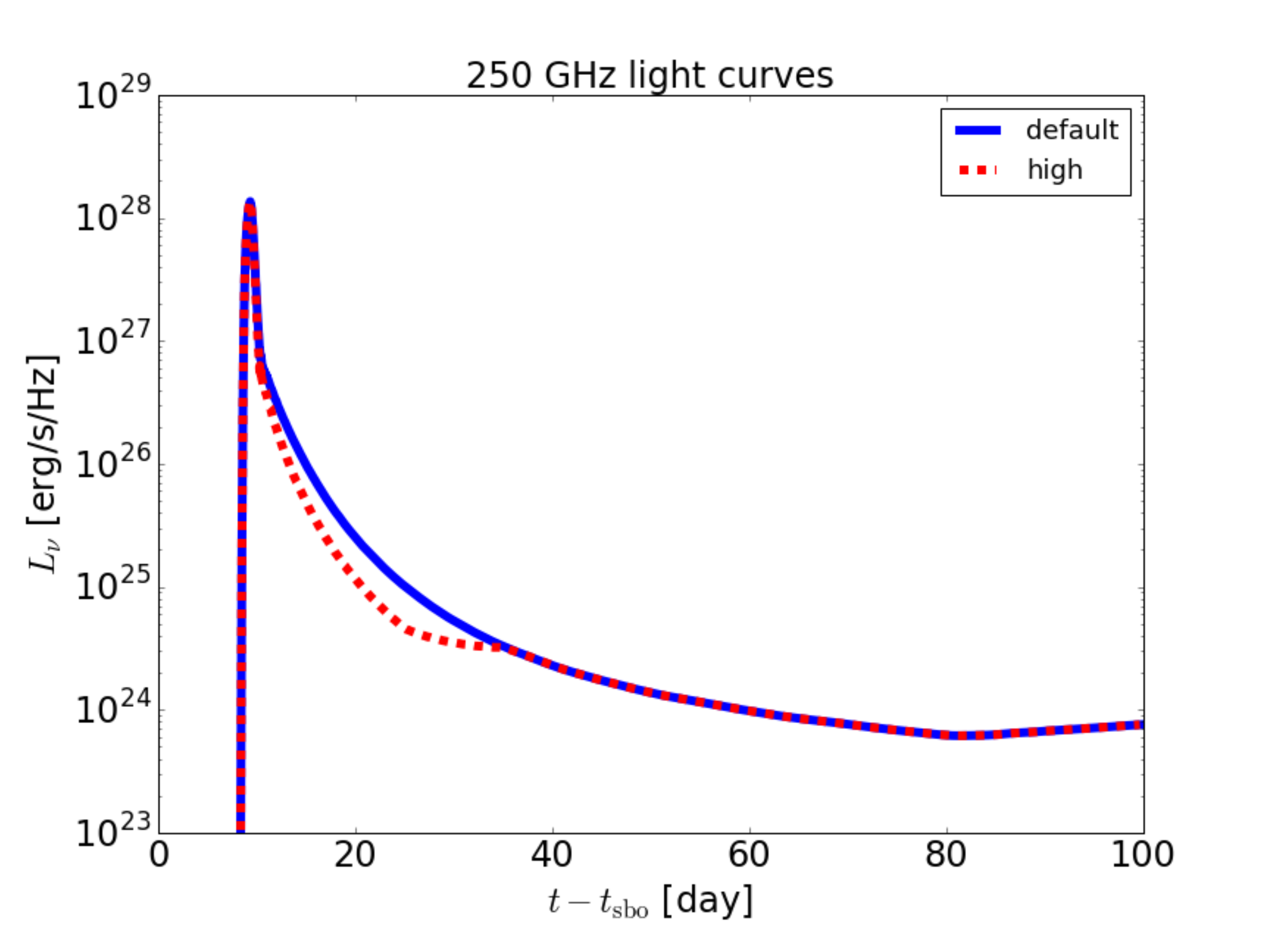}
\caption{The simulated 250 GHz light curves for `Mdot-2confined' model, with different choices of the bolometric light curve (Figure 3). Initially after the shock emerges out of the `confined' CSM ($t\sim 10 - 30$ days), the IC scattering dominates the cooling process and alters the radio light curves slightly. The peak behaviors, on the other hand, are hardly affected.}
\label{fig:LC_Lbol}
\end{figure}

\subsection{Observational prospects}
According to our results, the millimeter signals from an SN at a distance of several ten Mpc should be detectable with the observed flux density of $\mathcal{O}(1)$ mJy. We propose that this is an interesting target for ALMA, to robustly probe the existence of the `confined' CSM. Figure 9 shows the time sequence of the simulated synchrotron spectral energy distribution (SED), assuming that the SN explodes at a distance of 30 Mpc. There is a distinct gap between the very early ($t\lesssim 10$ days) and the later phase ($t\gtrsim 10$ days), %
and the degree of spectral difference depends on the density of the `confined' CSM.
We can see that the `confined' CSM produces strong signals in the millimeter range in the first 10 days, while the signal is totally damped in the centimeter range.
\begin{figure*}[ht!]
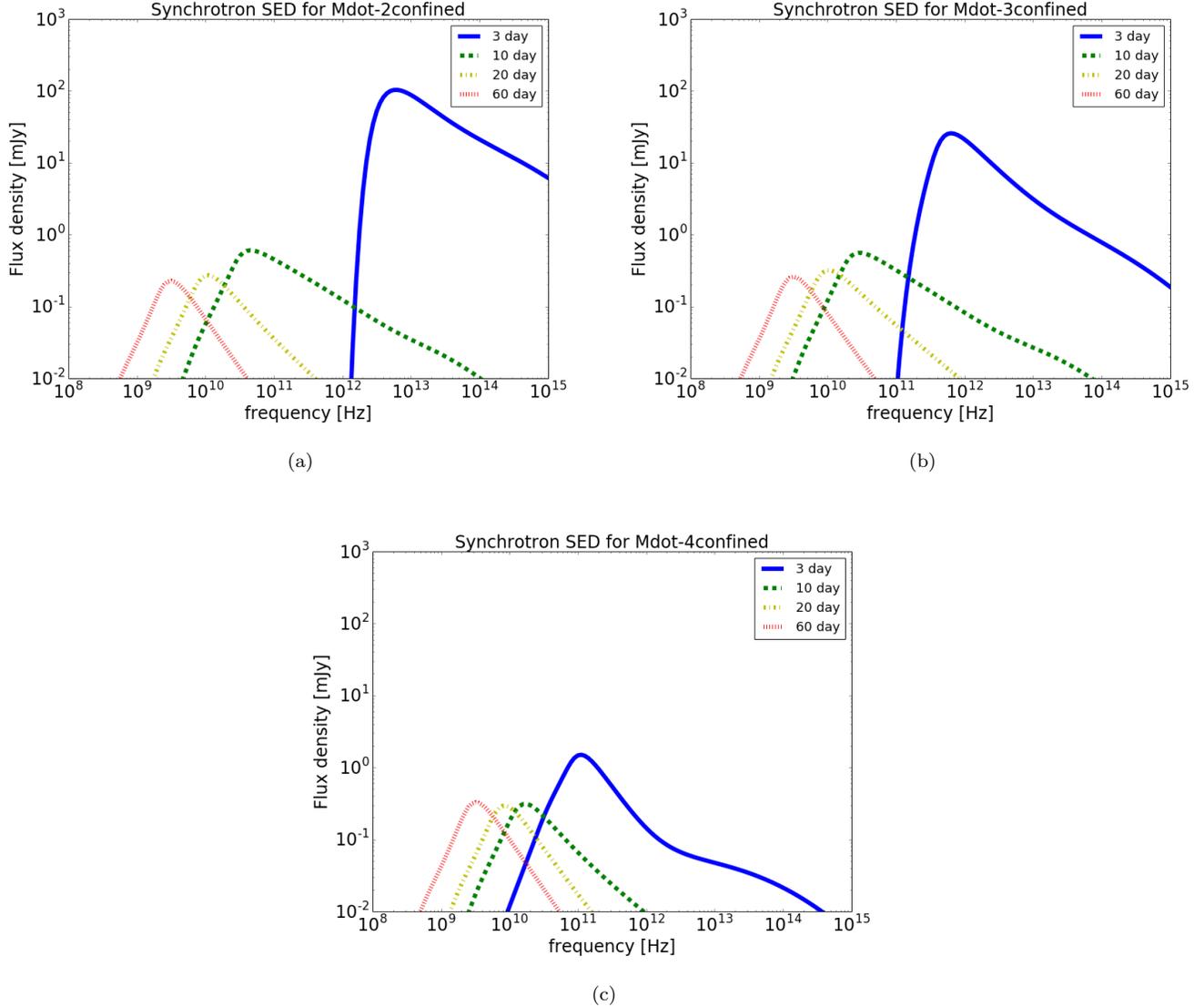

\gridline{
\fig{fig9_a.eps}{0.5\textwidth}{(a)}
\fig{fig9_b.eps}{0.5\textwidth}{(b)}
}
\gridline{
\fig{fig9_c.eps}{0.5\textwidth}{(c)}
}
\caption{The time series of the synchrotron emission SED for each model for an SN at 30 Mpc. Blue-solid, green-dashed, yellow-dash-dotted, and red-dotted lines show the SED at the phase $t = 5, 10, 20, 60$ days since the shock breakout, respectively. Each panel corresponds to the model (a) `Mdot-2confined', (b) `Mdot-3confined', (c) `Mdot-4confined'.}
\label{fig:synchrotronSED}
\end{figure*}

In practice when fitting the observational data, more detailed simulations are required to test several uncertainties in the present models. Especially, accurately deriving the CSM density scale can be difficult, as this is coupled with the unknown parameters, e.g., $\epsilon_{\rm e}$ and $\epsilon_{\rm B}$, while this degeneracy can be partly reduced by considering the effects of cooling processes \citep{2012ApJ...758...81M}. We, however, note that deriving the spatial distribution of the CSM density (whether it is smooth or truncated) is decoupled to these uncertainties, and can be robustly determined.

The multi-dimensional geometry of the CSM might also have an effect on the observed properties of the non-thermal emission, depending on the viewing angle. Observations indicating a global asymmetry in the CSM have indeed been reported for some SNe \citep{2000ApJ...536..239L, 2008ApJ...688.1186H, 2011A&A...527L...6P, 2016ApJ...832..194K}. 
Non-thermal emission modeling based on multi-dimensional hydrodynamic simulations will thus be of interest. We postpone such investigations to future works.

Finally, it is worth mentioning that while the electron spectral index has been constrained by observations so far ($\alpha_{e^-}=3$), the proton spectral index still involves uncertainties. The value we adopted ($\alpha_p = 3$ in our models) is different from the canonical value of 2 predicted by the test particle theory of DSA for strong non-relativistic shock \citep{1978MNRAS.182..147B, PhysRevLett.114.085003}. It is important because the proton spectral distribution is critical for secondary particle production. However, our models predict that if a `confined' CSM is present, the secondary particles dominate the synchrotron emission in the later phase ($t \gtrsim 10$ days). Therefore, future radio follow-up observations by ALMA will constrain the proton index which hopefully leads to the better understanding of particle acceleration at the SN shock.

\subsection{summary}

Thanks to recent high-cadence transient surveys and rapid follow-up observations, it has been revealed that some massive stars may release a large amount of their own mass within decades before the SN and form a `confined' CSM in the vicinity of the progenitors. However, previous investigations based on only optical data involve uncertainties in their interpretations. We suggest that radio synchrotron emission can alternatively be a more robust tracer of the `confined' CSM.

Our calculations basically follow general formalisms developed by previous works on the non-thermal emission from SNe; capturing the propagation of the shockwave, solving the particle energy distribution in the post-shocked CSM region, and estimating the synchrotron intensity. We perform spherically symmetric hydrodynamics simulations. Within the first 10 days, the forward shock propagates within the `confined' CSM. In the shocked CSM region, relativistic particles are injected via the particle acceleration at the shock through mechanisms such as DSA or the inelastic collisions of protons. Depending on the degree of the absorption (which depends on the CSM density), the emerging millimeter emission can robustly probe the existence of a `confined' CSM and its density (and spatial extent).

We have shown that strong millimeter emission is expected within the first 10 days since the shock breakout, and that the density of the `confined' CSM alters the behavior of the radio emission. The centimeter emission will be simply damped due to SSA and FFA. On the other hand, the millimeter emission is still detectable thanks to its transparencty, and thus it is a robust tracer of the confined CSM. In addition, we have shown that the peak luminosity is dominated by the primary electrons, accelerated by the shockwave, while in the later phase ($t\gtrsim 10$ days) the secondary electrons and positrons make a large contribution to the synchrotron radiation if the density of the `confined' CSM is larger than $\rho \gtrsim 10^{-15} 
\left(\frac{\dot{M}}{10^{-3}\ M_\odot\ \text{yr}^{-1}}\right)
\left(\frac{\dot{u_{\rm w}}}{100\ \text{km s}^{-1}}\right)^{-1}
\left(\frac{\dot{R_{\rm CSM}}}{7\times 10^{14} \ \text{cm}^{-2}}\right)^{-2}$ (corresponding to `Mdot-2confined' and `Mdot-3confined'). In summary, we propose that a target of opportunity observation by ALMA can provide strong diagnostics on the existence of the `confined' CSM and its nature.

\acknowledgments

The authors thank Akihiro Suzuki, Takashi Nagao, Ryo Sawada and Ryoma Ouchi for fruitful discussions. K.M. acknowledges support by JSPS KAKENHI Grant (18H04585, 18H05223, 17H02864). S.H.L. is supported by the Kyoto University Foundation and JSPS KAKENHI Grant Number JP19K03913.

\software{MESA \citep{2011ApJS..192....3P,2013ApJS..208....4P,2015ApJS..220...15P,2018ApJS..234...34P},\ 
SNEC \citep{2015ApJ...814...63M}}

\appendix
\section{Cooling Processes}
We include the synchrotron loss, inverse Compton scattering, adiabatic cooling, Coulomb cooling \citep{1979ApJ...227..364R,2002ApJ...571..866U}, and p-p collisions as the cooling processes \citep{2016MNRAS.460...44P}. The energy loss timescales are given as follows;
\begin{eqnarray}
t_{\rm i, syn} &=& \frac{6\pi m_{\rm e} c}{\sigma_{\rm T}}B(m_r)^{-2}\gamma_{\rm i}^{-1},\\
t_{\rm i, IC} &=& \frac{3m_{\rm e}c}{4\sigma_{\rm T}}U_{\rm ph}\gamma_{\rm i}^{-1},\\
t_{\rm i, ad} &=& \frac{3\gamma_{\rm i}}{\gamma_{\rm i}^2 -1}\frac{\mathcal{V}(m_r)}{\dot{\mathcal{V}(m_r)}},\\
t_{\rm i, coulomb} &=& \frac{2E_{\rm i} m_{\rm i} c}{8\pi q^4 n_{\rm H} \ln\Lambda}, \\
t_{\rm pp} &=& (\kappa_{\rm pp}\sigma_{\rm pp} n_{\rm H} c)^{-1},
\end{eqnarray}
where $\sigma_{\rm T}, \gamma_{\rm i}, U_{\rm ph}, \dot{\mathcal{V}}, n_{\rm H}, \ln\Lambda, \kappa_{\rm pp}=0.5$ and $\sigma_{\rm pp}$ are the Thomson scattering cross section, Lorentz factor of the particle $\gamma_{\rm i} = E_{\rm i}/m_{\rm i}c^2$, the energy density of the photon, the time derivative of the volume of the zone at the radius $r$, the number density of the thermal gas, the Coulomb logarithm, the inelasticity of the proton and the total inelastic cross section of proton, respectively. In this study the coulomb logarithm is fixed to be 30. $U_{\rm ph}$ is determined by the SN bolometric luminosity as
\begin{eqnarray}
U_{\rm ph} = \frac{L_{\rm bol}}{\pi r^2 c}.
\end{eqnarray}
For the proton inelastic energy loss cross section, we employ the formula developed by \cite{2006PhRvD..74c4018K},
\begin{eqnarray}
\sigma_{\rm pp} = (34.3 + 1.88L + 0.25L^2) \times(1-(E_{\rm th}/E_{\rm p})^4)^2 \ \ 10^{-27} \rm{cm}^{2},
\end{eqnarray}
where $L = \ln(E_{\rm p}/1\mbox{ TeV})$, and $E_{\rm th} = 1.22 \mbox{ GeV}$ which is the threshold of proton energy in order for generating pions.

The total cooling timescales ($t_{\rm i, loss}$) for each particles are defined as follows;
\begin{eqnarray}
\frac{1}{t_{\rm i, loss}} = \frac{1}{t_{\rm i, syn}} & + &\frac{1}{t_{\rm i, IC}} + \frac{1}{t_{\rm i, coulomb}} + \frac{1}{t_{\rm i, adiabatic}} \ ({\rm i} = e^-, e^+), \\
\frac{1}{t_{\rm p, loss}} & = &\frac{1}{t_{\rm pp}} + \frac{1}{t_{\rm p, adiabatic}} + \frac{1}{t_{\rm p, coulomb}}.
\end{eqnarray}

\section{Synchrotron emission}
In this section we describe the formalization of the synchrotron radiation used in our simulations \citep{1979rpa..book.....R}. With the critical frequency $\nu_c$, the emission power per unit frequency is given by
\begin{eqnarray}
P_\nu(\gamma) &=& \frac{\sqrt{3}q^3 B \sin \theta}{m_{\rm e} c^2} F
\left(
\frac{\nu}{\nu_{\rm c}}
\right), \\
\nu_{\rm c} &=& \frac{3\gamma^2 q B \sin\theta}{4\pi m_{\rm e} c},
\end{eqnarray}
where $\theta$ is the pitch angle. The function $F(x)$ is defined by the following function,
\begin{eqnarray}
F(x) = x\int_x^\infty K_{5/3}(y)dy,
\end{eqnarray}
where $K_{5/3}(y)$ is the modified Bessel function of 5/3 order.
In the context of shock acceleration, we assume that the magnetic field is amplified as turbulence and the pitch angle is distributed isotropically. We thus employ the angle-averaged formula, 
\begin{eqnarray}
\bar{P}_\nu(\gamma) = \frac{\sqrt{3}q^3 B}{m_{\rm e}c^2} G
&\left(
{\displaystyle \frac{\nu}{\nu_{\rm c}'}}
\right)&, \ \nu_{\rm c}' = \frac{3\gamma^2qB}{4\pi m_{\rm e} c},\\
G(x) = \int \sin\theta F\left(\frac{x}{\sin\theta}\right) &{\displaystyle \frac{d\Omega}{4\pi}}& = x\int_x^\infty K_{5/3}(\xi)\sqrt{1-\frac{x^2}{\xi^2}} d\xi.
\end{eqnarray}
While $G(x)$ can be expressed by the modified Bessel function, we instead adopt the following approximated formulae \citep{2010PhRvD..82d3002A};
\begin{eqnarray}
G(x) \simeq \frac{1.808x^{1/3}}{\sqrt{1+3.4x^{2/3}}}\frac{1+2.21x^{2/3}+0.347x^{4/3}}{1+1.353x^{2/3}+0.217x^{4/3}}e^{-x}.
\end{eqnarray}

\bibliography{manuscript}

\begin{thebibliography}{}
\expandafter\ifx\csname natexlab\endcsname\relax\def\natexlab#1{#1}\fi

\bibitem[{{Aharonian} {et~al.}(2010){Aharonian}, {Kelner}, \&
  {Prosekin}}]{2010PhRvD..82d3002A}
{Aharonian}, F.~A., {Kelner}, S.~R., \& {Prosekin}, A.~Y. 2010, \prd, 82,
  043002

\bibitem[{{Anderson} {et~al.}(2014){Anderson}, {Gonz{\'a}lez-Gait{\'a}n},
  {Hamuy}, {Guti{\'e}rrez}, {Stritzinger}, {Olivares E.}, {Phillips},
  {Schulze}, {Antezana}, {Bolt}, {Campillay}, {Castell{\'o}n}, {Contreras}, {de
  Jaeger}, {Folatelli}, {F{\"o}rster}, {Freedman}, {Gonz{\'a}lez}, {Hsiao},
  {Krzemi{\'n}ski}, {Krisciunas}, {Maza}, {McCarthy}, {Morrell}, {Persson},
  {Roth}, {Salgado}, {Suntzeff}, \& {Thomas-Osip}}]{2014ApJ...786...67A}
{Anderson}, J.~P., {Gonz{\'a}lez-Gait{\'a}n}, S., {Hamuy}, M., {et~al.} 2014,
  \apj, 786, 67

\bibitem[{{Bell}(1978)}]{1978MNRAS.182..147B}
{Bell}, A.~R. 1978, \mnras, 182, 147

\bibitem[{{Bellm} {et~al.}(2019){Bellm}, {Kulkarni}, {Graham}, {Dekany},
  {Smith}, {Riddle}, {Masci}, {Helou}, {Prince}, {Adams}, {Barbarino},
  {Barlow}, {Bauer}, {Beck}, {Belicki}, {Biswas}, {Blagorodnova}, {Bodewits},
  {Bolin}, {Brinnel}, {Brooke}, {Bue}, {Bulla}, {Burruss}, {Cenko}, {Chang},
  {Connolly}, {Coughlin}, {Cromer}, {Cunningham}, {De}, {Delacroix}, {Desai},
  {Duev}, {Eadie}, {Farnham}, {Feeney}, {Feindt}, {Flynn}, {Franckowiak},
  {Frederick}, {Fremling}, {Gal-Yam}, {Gezari}, {Giomi}, {Goldstein},
  {Golkhou}, {Goobar}, {Groom}, {Hacopians}, {Hale}, {Henning}, {Ho}, {Hover},
  {Howell}, {Hung}, {Huppenkothen}, {Imel}, {Ip}, {Ivezi{\'c}}, {Jackson},
  {Jones}, {Juric}, {Kasliwal}, {Kaspi}, {Kaye}, {Kelley}, {Kowalski},
  {Kramer}, {Kupfer}, {Landry}, {Laher}, {Lee}, {Lin}, {Lin}, {Lunnan},
  {Giomi}, {Mahabal}, {Mao}, {Miller}, {Monkewitz}, {Murphy}, {Ngeow},
  {Nordin}, {Nugent}, {Ofek}, {Patterson}, {Penprase}, {Porter}, {Rauch},
  {Rebbapragada}, {Reiley}, {Rigault}, {Rodriguez}, {van Roestel}, {Rusholme},
  {van Santen}, {Schulze}, {Shupe}, {Singer}, {Soumagnac}, {Stein}, {Surace},
  {Sollerman}, {Szkody}, {Taddia}, {Terek}, {Van Sistine}, {van Velzen},
  {Vestrand}, {Walters}, {Ward}, {Ye}, {Yu}, {Yan}, \&
  {Zolkower}}]{2019PASP..131a8002B}
{Bellm}, E.~C., {Kulkarni}, S.~R., {Graham}, M.~J., {et~al.} 2019, \pasp, 131,
  018002

\bibitem[{{Caprioli} {et~al.}(2009){Caprioli}, {Blasi}, \&
  {Amato}}]{2009MNRAS.396.2065C}
{Caprioli}, D., {Blasi}, P., \& {Amato}, E. 2009, Monthly Notices of the Royal
  Astronomical Society, 396, 2065

\bibitem[{{Caprioli} {et~al.}(2015){Caprioli}, {Pop}, \&
  {Spitkovsky}}]{2015ApJ...798L..28C}
{Caprioli}, D., {Pop}, A.-R., \& {Spitkovsky}, A. 2015, \apjl, 798, L28

\bibitem[{{Chevalier}(1982{\natexlab{a}})}]{1982ApJ...258..790C}
{Chevalier}, R.~A. 1982{\natexlab{a}}, \apj, 258, 790

\bibitem[{{Chevalier}(1982{\natexlab{b}})}]{1982ApJ...259..302C}
---. 1982{\natexlab{b}}, \apj, 259, 302

\bibitem[{{Chevalier}(1998)}]{1998ApJ...499..810C}
---. 1998, \apj, 499, 810

\bibitem[{{Chevalier} \& {Fransson}(2006)}]{2006ApJ...651..381C}
{Chevalier}, R.~A., \& {Fransson}, C. 2006, \apj, 651, 381

\bibitem[{{Chevalier} \& {Fransson}(2017)}]{2017hsn..book..875C}
---. 2017, {Thermal and Non-thermal Emission from Circumstellar Interaction},
  ed. A.~W. {Alsabti} \& P.~{Murdin}, 875

\bibitem[{{Chevalier} {et~al.}(2006){Chevalier}, {Fransson}, \&
  {Nymark}}]{2006ApJ...641.1029C}
{Chevalier}, R.~A., {Fransson}, C., \& {Nymark}, T.~K. 2006, \apj, 641, 1029

\bibitem[{{Drury}(1983)}]{1983RPPh...46..973D}
{Drury}, L.~O. 1983, Reports on Progress in Physics, 46, 973

\bibitem[{{Fermi}(1949)}]{1949PhRv...75.1169F}
{Fermi}, E. 1949, Physical Review, 75, 1169

\bibitem[{{Filippenko}(1997)}]{1997ARA&A..35..309F}
{Filippenko}, A.~V. 1997, \araa, 35, 309

\bibitem[{{Forster} {et~al.}(2018){Forster}, {Moriya}, {Maureira}, {Anderson},
  {Blinnikov}, {Bufano}, {Cabrera-Vives}, {Clocchiatti}, {de Jaeger},
  {Estevez}, {Galbany}, {Gonzalez-Gaitan}, {Grafener}, {Hamuy}, {Hsiao},
  {Huentelemu}, {Huijse}, {Kuncarayakti}, {Martinez}, {Medina}, {Olivares},
  {Pignata}, {Razza}, {Reyes}, {San}, {Smith}, {Vera}, {Vivas}, {de Ugarte
  Postigo}, {Yoon}, {Ashall}, {Fraser}, {Gal-Yam}, {Kankare}, {Le Guillou},
  {Mazzali}, {Walton}, \& {Young}}]{2018NatAs...2..808F}
{Forster}, F., {Moriya}, T.~J., {Maureira}, J.~C., {et~al.} 2018, Nature
  Astronomy, 2, 808

\bibitem[{{Fransson} \& {Bj{\"o}rnsson}(1998)}]{1998ApJ...509..861F}
{Fransson}, C., \& {Bj{\"o}rnsson}, C.-I. 1998, \apj, 509, 861

\bibitem[{{Fuller}(2017)}]{2017MNRAS.470.1642F}
{Fuller}, J. 2017, \mnras, 470, 1642

\bibitem[{{Gal-Yam} {et~al.}(2014){Gal-Yam}, {Arcavi}, {Ofek}, {Ben-Ami},
  {Cenko}, {Kasliwal}, {Cao}, {Yaron}, {Tal}, {Silverman}, {Horesh}, {De Cia},
  {Taddia}, {Sollerman}, {Perley}, {Vreeswijk}, {Kulkarni}, {Nugent},
  {Filippenko}, \& {Wheeler}}]{2014Natur.509..471G}
{Gal-Yam}, A., {Arcavi}, I., {Ofek}, E.~O., {et~al.} 2014, \nat, 509, 471

\bibitem[{{Graham} {et~al.}(2019){Graham}, {Kulkarni}, {Bellm}, {Adams},
  {Barbarino}, {Blagorodnova}, {Bodewits}, {Bolin}, {Brady}, {Cenko}, {Chang},
  {Coughlin}, {De}, {Eadie}, {Farnham}, {Feindt}, {Franckowiak}, {Fremling},
  {Gal-yam}, {Gezari}, {Ghosh}, {Goldstein}, {Golkhou}, {Goobar}, {Ho},
  {Huppenkothen}, {Ivezic}, {Jones}, {Juric}, {Kaplan}, {Kasliwal}, {Kelley},
  {Kupfer}, {Lee}, {Lin}, {Lunnan}, {Mahabal}, {Miller}, {Ngeow}, {Nugent},
  {Ofek}, {Prince}, {Rauch}, {van Roestel}, {Schulze}, {Singer}, {Sollerman},
  {Taddia}, {Yan}, {Ye}, {Yu}, {Andreoni}, {Barlow}, {Bauer}, {Beck},
  {Belicki}, {Biswas}, {Brinnel}, {Brooke}, {Bue}, {Bulla}, {Burdge},
  {Burruss}, {Connolly}, {Cromer}, {Cunningham}, {Dekany}, {Delacroix},
  {Desai}, {Duev}, {Hacopians}, {Hale}, {Helou}, {Henning}, {Hover},
  {Hillenbrand}, {Howell}, {Hung}, {Imel}, {Ip}, {Jackson}, {Kaspi}, {Kaye},
  {Kowalski}, {Kramer}, {Kuhn}, {Land ry}, {Laher}, {Mao}, {Masci},
  {Monkewitz}, {Murphy}, {Nordin}, {Patterson}, {Penprase}, {Porter},
  {Rebbapragada}, {Reiley}, {Riddle}, {Rigault}, {Rodriguez}, {Rusholme}, {van
  Santen}, {Shupe}, {Smith}, {Soumagnac}, {Stein}, {Surace}, {Szkody}, {Terek},
  {van Sistine}, {van Velzen}, {Vestrand}, {Walters}, {Ward}, {Zhang}, \&
  {Zolkower}}]{2019arXiv190201945G}
{Graham}, M.~J., {Kulkarni}, S.~R., {Bellm}, E.~C., {et~al.} 2019, arXiv
  e-prints, arXiv:1902.01945

\bibitem[{{Groh}(2014)}]{2014A&A...572L..11G}
{Groh}, J.~H. 2014, \aap, 572, L11

\bibitem[{{Hoffman} {et~al.}(2008){Hoffman}, {Leonard}, {Chornock},
  {Filippenko}, {Barth}, \& {Matheson}}]{2008ApJ...688.1186H}
{Hoffman}, J.~L., {Leonard}, D.~C., {Chornock}, R., {et~al.} 2008, \apj, 688,
  1186

\bibitem[{{Kamae} {et~al.}(2006){Kamae}, {Karlsson}, {Mizuno}, {Abe}, \&
  {Koi}}]{2006ApJ...647..692K}
{Kamae}, T., {Karlsson}, N., {Mizuno}, T., {Abe}, T., \& {Koi}, T. 2006, \apj,
  647, 692

\bibitem[{{Katsuda} {et~al.}(2016){Katsuda}, {Maeda}, {Bamba}, {Terada},
  {Fukazawa}, {Kawabata}, {Ohno}, {Sugawara}, {Tsuboi}, \&
  {Immler}}]{2016ApJ...832..194K}
{Katsuda}, S., {Maeda}, K., {Bamba}, A., {et~al.} 2016, \apj, 832, 194

\bibitem[{{Kelner} {et~al.}(2006){Kelner}, {Aharonian}, \&
  {Bugayov}}]{2006PhRvD..74c4018K}
{Kelner}, S.~R., {Aharonian}, F.~A., \& {Bugayov}, V.~V. 2006, \prd, 74, 034018

\bibitem[{{Khazov} {et~al.}(2016){Khazov}, {Yaron}, {Gal-Yam}, {Manulis},
  {Rubin}, {Kulkarni}, {Arcavi}, {Kasliwal}, {Ofek}, {Cao}, {Perley},
  {Sollerman}, {Horesh}, {Sullivan}, {Filippenko}, {Nugent}, {Howell}, {Cenko},
  {Silverman}, {Ebeling}, {Taddia}, {Johansson}, {Laher}, {Surace},
  {Rebbapragada}, {Wozniak}, \& {Matheson}}]{2016ApJ...818....3K}
{Khazov}, D., {Yaron}, O., {Gal-Yam}, A., {et~al.} 2016, \apj, 818, 3

\bibitem[{{Langer}(2012)}]{2012ARA&A..50..107L}
{Langer}, N. 2012, \araa, 50, 107

\bibitem[{{Law} {et~al.}(2009){Law}, {Kulkarni}, {Dekany}, {Ofek}, {Quimby},
  {Nugent}, {Surace}, {Grillmair}, {Bloom}, {Kasliwal}, {Bildsten}, {Brown},
  {Cenko}, {Ciardi}, {Croner}, {Djorgovski}, {van Eyken}, {Filippenko}, {Fox},
  {Gal-Yam}, {Hale}, {Hamam}, {Helou}, {Henning}, {Howell}, {Jacobsen},
  {Laher}, {Mattingly}, {McKenna}, {Pickles}, {Poznanski}, {Rahmer}, {Rau},
  {Rosing}, {Shara}, {Smith}, {Starr}, {Sullivan}, {Velur}, {Walters}, \&
  {Zolkower}}]{2009PASP..121.1395L}
{Law}, N.~M., {Kulkarni}, S.~R., {Dekany}, R.~G., {et~al.} 2009, \pasp, 121,
  1395

\bibitem[{{Lee} {et~al.}(2012){Lee}, {Ellison}, \&
  {Nagataki}}]{2012ApJ...750..156L}
{Lee}, S.-H., {Ellison}, D.~C., \& {Nagataki}, S. 2012, \apj, 750, 156

\bibitem[{{Lee} {et~al.}(2013){Lee}, {Slane}, {Ellison}, {Nagataki}, \&
  {Patnaude}}]{2013ApJ...767...20L}
{Lee}, S.-H., {Slane}, P.~O., {Ellison}, D.~C., {Nagataki}, S., \& {Patnaude},
  D.~J. 2013, \apj, 767, 20

\bibitem[{{Leonard} {et~al.}(2000){Leonard}, {Filippenko}, {Barth}, \&
  {Matheson}}]{2000ApJ...536..239L}
{Leonard}, D.~C., {Filippenko}, A.~V., {Barth}, A.~J., \& {Matheson}, T. 2000,
  \apj, 536, 239

\bibitem[{{Maeda}(2012)}]{2012ApJ...758...81M}
{Maeda}, K. 2012, \apj, 758, 81

\bibitem[{{Maeda}(2013)}]{2013ApJ...762...14M}
---. 2013, \apj, 762, 14

\bibitem[{{Moriya} {et~al.}(2017){Moriya}, {Yoon}, {Gr{\"a}fener}, \&
  {Blinnikov}}]{2017MNRAS.469L.108M}
{Moriya}, T.~J., {Yoon}, S.-C., {Gr{\"a}fener}, G., \& {Blinnikov}, S.~I. 2017,
  \mnras, 469, L108

\bibitem[{{Morozova} {et~al.}(2015){Morozova}, {Piro}, {Renzo}, {Ott},
  {Clausen}, {Couch}, {Ellis}, \& {Roberts}}]{2015ApJ...814...63M}
{Morozova}, V., {Piro}, A.~L., {Renzo}, M., {et~al.} 2015, \apj, 814, 63

\bibitem[{{Morozova} {et~al.}(2017){Morozova}, {Piro}, \&
  {Valenti}}]{2017ApJ...838...28M}
{Morozova}, V., {Piro}, A.~L., \& {Valenti}, S. 2017, \apj, 838, 28

\bibitem[{{Murase} {et~al.}(2019){Murase}, {Franckowiak}, {Maeda}, {Margutti},
  \& {Beacom}}]{2019ApJ...874...80M}
{Murase}, K., {Franckowiak}, A., {Maeda}, K., {Margutti}, R., \& {Beacom},
  J.~F. 2019, \apj, 874, 80

\bibitem[{{Ouchi} \& {Maeda}(2019)}]{2019arXiv190407878O}
{Ouchi}, R., \& {Maeda}, K. 2019, arXiv e-prints, arXiv:1904.07878

\bibitem[{Park {et~al.}(2015)Park, Caprioli, \&
  Spitkovsky}]{PhysRevLett.114.085003}
Park, J., Caprioli, D., \& Spitkovsky, A. 2015, Phys. Rev. Lett., 114, 085003

\bibitem[{{Patat} {et~al.}(2011){Patat}, {Taubenberger}, {Benetti},
  {Pastorello}, \& {Harutyunyan}}]{2011A&A...527L...6P}
{Patat}, F., {Taubenberger}, S., {Benetti}, S., {Pastorello}, A., \&
  {Harutyunyan}, A. 2011, \aap, 527, L6

\bibitem[{{Paxton} {et~al.}(2011){Paxton}, {Bildsten}, {Dotter}, {Herwig},
  {Lesaffre}, \& {Timmes}}]{2011ApJS..192....3P}
{Paxton}, B., {Bildsten}, L., {Dotter}, A., {et~al.} 2011, \apjs, 192, 3

\bibitem[{{Paxton} {et~al.}(2013){Paxton}, {Cantiello}, {Arras}, {Bildsten},
  {Brown}, {Dotter}, {Mankovich}, {Montgomery}, {Stello}, {Timmes}, \&
  {Townsend}}]{2013ApJS..208....4P}
{Paxton}, B., {Cantiello}, M., {Arras}, P., {et~al.} 2013, \apjs, 208, 4

\bibitem[{{Paxton} {et~al.}(2015){Paxton}, {Marchant}, {Schwab}, {Bauer},
  {Bildsten}, {Cantiello}, {Dessart}, {Farmer}, {Hu}, {Langer}, {Townsend},
  {Townsley}, \& {Timmes}}]{2015ApJS..220...15P}
{Paxton}, B., {Marchant}, P., {Schwab}, J., {et~al.} 2015, \apjs, 220, 15

\bibitem[{{Paxton} {et~al.}(2018){Paxton}, {Schwab}, {Bauer}, {Bildsten},
  {Blinnikov}, {Duffell}, {Farmer}, {Goldberg}, {Marchant}, {Sorokina},
  {Thoul}, {Townsend}, \& {Timmes}}]{2018ApJS..234...34P}
{Paxton}, B., {Schwab}, J., {Bauer}, E.~B., {et~al.} 2018, \apjs, 234, 34

\bibitem[{{Petropoulou} {et~al.}(2016){Petropoulou}, {Kamble}, \&
  {Sironi}}]{2016MNRAS.460...44P}
{Petropoulou}, M., {Kamble}, A., \& {Sironi}, L. 2016, \mnras, 460, 44

\bibitem[{{Rau} {et~al.}(2009){Rau}, {Kulkarni}, {Law}, {Bloom}, {Ciardi},
  {Djorgovski}, {Fox}, {Gal-Yam}, {Grillmair}, {Kasliwal}, {Nugent}, {Ofek},
  {Quimby}, {Reach}, {Shara}, {Bildsten}, {Cenko}, {Drake}, {Filippenko},
  {Helfand}, {Helou}, {Howell}, {Poznanski}, \&
  {Sullivan}}]{2009PASP..121.1334R}
{Rau}, A., {Kulkarni}, S.~R., {Law}, N.~M., {et~al.} 2009, \pasp, 121, 1334

\bibitem[{{Rephaeli}(1979)}]{1979ApJ...227..364R}
{Rephaeli}, Y. 1979, \apj, 227, 364

\bibitem[{{Rybicki} \& {Lightman}(1979)}]{1979rpa..book.....R}
{Rybicki}, G.~B., \& {Lightman}, A.~P. 1979, {Radiative processes in
  astrophysics}

\bibitem[{{Slane} {et~al.}(2014){Slane}, {Lee}, {Ellison}, {Patnaude},
  {Hughes}, {Eriksen}, {Castro}, \& {Nagataki}}]{2014ApJ...783...33S}
{Slane}, P., {Lee}, S.~H., {Ellison}, D.~C., {et~al.} 2014, \apj, 783, 33

\bibitem[{{Smartt}(2015)}]{2015PASA...32...16S}
{Smartt}, S.~J. 2015, \pasa, 32, e016

\bibitem[{{Uchiyama} {et~al.}(2002){Uchiyama}, {Takahashi}, {Aharonian}, \&
  {Mattox}}]{2002ApJ...571..866U}
{Uchiyama}, Y., {Takahashi}, T., {Aharonian}, F.~A., \& {Mattox}, J.~R. 2002,
  \apj, 571, 866

\bibitem[{{Yaron} {et~al.}(2017){Yaron}, {Perley}, {Gal-Yam}, {Groh}, {Horesh},
  {Ofek}, {Kulkarni}, {Sollerman}, {Fransson}, {Rubin}, {Szabo}, {Sapir},
  {Taddia}, {Cenko}, {Valenti}, {Arcavi}, {Howell}, {Kasliwal}, {Vreeswijk},
  {Khazov}, {Fox}, {Cao}, {Gnat}, {Kelly}, {Nugent}, {Filippenko}, {Laher},
  {Wozniak}, {Lee}, {Rebbapragada}, {Maguire}, {Sullivan}, \&
  {Soumagnac}}]{2017NatPh..13..510Y}
{Yaron}, O., {Perley}, D.~A., {Gal-Yam}, A., {et~al.} 2017, Nature Physics, 13,
  510

\bibitem[{{Yasuda} \& {Lee}(2019)}]{2019ApJ...876...27Y}
{Yasuda}, H., \& {Lee}, S.-H. 2019, \apj, 876, 27

\end{thebibliography}
\end{document}